\documentclass[a4paper]{article}

\usepackage{JCMS}           
\usepackage[utf8]{inputenc} 
\usepackage[T1]{fontenc}    
\usepackage{type1cm}        %
\usepackage{booktabs}       
\usepackage{multirow}       
\usepackage{array}          
\usepackage{placeins}       
\usepackage{amsmath}        
\usepackage{amsfonts}       
\usepackage{amssymb}        
\usepackage{amsthm}         
\usepackage{nicefrac}       
\usepackage{microtype}      
\usepackage{graphicx}       

\usepackage{tikz}
\usetikzlibrary{arrows.meta}
\usetikzlibrary{shapes.geometric}
\usepackage{pgfplots}
\pgfplotsset{compat=1.18}

\usepackage{algorithm}
\usepackage{algpseudocode}

\usepackage{csquotes}       
\usepackage[backend=biber,style=apa,natbib=true]{biblatex} 
\usepackage{url}            
\usepackage{hyperref}       
\hypersetup{pdftitle={Amanous: Distribution-Switching for Superhuman Piano Density on Disklavier}, pdfauthor={Joonhyung Bae}}

\newcommand{\tbl}[2]{\caption{#1}\centering#2}
\newcommand{\tabnote}[1]{\par\vspace{0.5ex}\begin{minipage}{\linewidth}\footnotesize\raggedright#1\end{minipage}}

\theoremstyle{definition}
\newtheorem{definition}{Definition}
\theoremstyle{remark}

\title{Amanous: Distribution-Switching for Superhuman Piano Density on Disklavier}

\author{%
  \textbf{Joonhyung Bae} \\
  Culture Technology Research Institute \\
  KAIST \\
  Daejeon, Republic of Korea \\
  \texttt{jh.bae@kaist.ac.kr} \\
}

\begin{document}

\maketitle

\begin{abstract}
A player piano can strike more keys, across more of the register, and in faster succession than any pianist can reach. Composers have pursued that possibility for decades, and three traditions dominate the attempt, namely Nancarrow's tempo canons, Xenakis' stochastic distributions, and L-system grammars. These traditions have developed in isolation, on abstractions that do not obviously combine, and none of them accounts for the instrument itself. A Disklavier does not answer instantly, because a loud note reaches the string sooner than a soft one, so music written as if the mechanism were transparent arrives at the listener distorted. We present \emph{Amanous}, a hardware-aware composition system for the Yamaha Disklavier that unifies the three traditions through \emph{distribution-switching}, in which each grammar symbol selects an entire distributional regime rather than adjusting parameters within a fixed one. A four-layer pipeline carries a symbol from grammar to actuation-ready MIDI. An L-system fixes the macro-form, each symbol is mapped to a set of distributions and a tempo-canon ratio, events are sampled and time-scaled, and a hardware abstraction layer pre-compensates the instrument's velocity-dependent latency before any MIDI is emitted. Tempo-canon convergence points close a feedback loop that lets the music's own temporal structure trigger the next distributional switch. We show that compositional intent survives this pipeline. Sections generated from different symbols remain statistically separable at the output. Ablating the L-system, the tempo canon, or the hardware compensation each degrades a distinct and measurable property of the result, and hardware compensation takes the mean onset error, under the modelled latency curve, from roughly 18~ms to below the instrument's own 1~ms scanning resolution. We further find that structured and random textures are most separable near 25~notes/s and that melodic metrics lose their discriminative power over the 40--100~notes/s band, above which the remaining difference is consistent with distributional content rather than melodic order. That band marks the density range beyond which cross-domain constraints become the effective compositional lever. All reported results are computational, nothing was played or measured on an instrument, and a psychoacoustic validation protocol is proposed for future work.\footnote{Supplementary materials (Excerpts~1--4) accompany this article as a compressed audio archive. Source code: \url{https://github.com/joonhyungbae/amanous}.}
\end{abstract}

\section{Introduction}
\label{sec:intro}

A piano has eighty-eight keys, and a pianist has ten fingers. Everything in the repertoire is shaped by that arithmetic, including what counts as a dense texture, a wide leap, or a fast passage. An automated piano removes the arithmetic. The Yamaha Disklavier accepts MIDI at rates that saturate its own mechanism, so it will play chords no hand can span and note rates no hand can sustain. This opens a region of musical possibility that has been visible for a century and inhabited by very few. Conlon Nancarrow spent decades there, punching paper rolls for a player piano because no performer could realise what he had written \citep{gann1995music}. What he could not have was control in the medium itself, since the roll fixed everything in advance and answered no questions about what it would sound like.

The words \emph{superhuman} and \emph{beyond-human} are used here in a strictly physical sense. They denote event densities, degrees of polyphony, and register spans that lie outside what a human pianist can execute, given a fixed number of fingers, a minimum achievable inter-onset interval (IOI) per hand, and a bounded reach across the keyboard. The sense is therefore mechanical rather than evaluative. It differs from the usage common in artificial-intelligence research, where a superhuman system is one that outperforms people at a task. Amanous is not measured against human composers or performers on any criterion of quality or skill. It operates in a region of the parameter space that human performance cannot physically enter, and the questions asked here concern what compositional structure can survive in that region.

Three traditions have supplied the tools for composing in this region, and each solved a different part of the problem. Nancarrow's tempo canons give large-scale temporal architecture, letting voices move at different rates and meet at moments the composer chooses \citep{gann1995music, nemire2014convergence}. Xenakis showed that a probability distribution can itself be a compositional object, so that a texture is specified by its statistics rather than note by note \citep{xenakis1992formalized}. L-system grammars generate self-similar form by recursive rewriting, producing hierarchical structure from a handful of rules \citep{prusinkiewicz1990algorithmic, manousakis2006musical}. Each tradition is mature, and each is normally used alone.

They are used alone because they do not obviously fit together. A tempo canon speaks in ratios, a stochastic process speaks in distributions, and a grammar speaks in symbols, and it is not clear what it would mean for a symbol to select a distribution that is then time-scaled by a ratio. Nor does any of the three take account of the instrument. A Disklavier's key is thrown by a solenoid, and the hammer's flight time depends on how hard the key is driven, so the delay between a MIDI command and the string varies with velocity. Measurements of a Disklavier grand report reproduction onset errors spanning $-20$ to $+30$~ms \citep{goebl2003measurement}, and no compositional framework has treated that latency as part of the composition. The result is a gap that widens exactly where the music gets interesting. The denser the texture, the more the three abstractions need to cooperate, and the more the mechanism's own timing intrudes.

There is a further reason the dense end of the range is hard to work in. Listeners parse sound into streams by pitch proximity and temporal continuity \citep{bregman1990auditory}, but that capacity has a limit, and past roughly twenty events per second the ear stops following individual lines and hears an aggregate texture instead \citep{roads2001microsound}. Where the limit falls is not fixed by rate alone, because whether a sequence coheres or splits also depends on the pitch interval between successive tones \citep{vannoorden1975temporal}. A composer writing above that limit is writing for a mode of listening that the tools were not designed for, and has no way of knowing where the boundary lies or which compositional levers still do any work beyond it.

Two terms from the tempo-canon literature recur throughout and are introduced here before they are used. A \emph{tempo canon} is a canon in which the voices state the same material at different rates, so that a voice sounding at a tempo ratio of 3:4 against another advances through the material three quarters as fast, and the voices drift steadily out of alignment rather than remaining in fixed rhythmic register. A \emph{convergence point} (CP) is a moment at which two such voices, drifting at different rates, coincide again in time. Nancarrow used convergence points as the formal pillars of his canons, timing them to fall at structurally decisive moments. Because a coincidence is exact only for rational tempo ratios, a computational treatment requires a tolerance parameter $\epsilon$ that defines how close two onsets must fall to count as convergent. Both terms are formalised in Equation~\ref{eq:canon} and Definition~\ref{def:cp}.

Our starting intuition is that the three traditions are not really competitors, because they operate at different time scales and can be made to hand work to one another. A grammar is good at deciding the order of large sections. A tempo canon is good at organising the middle scale, where voices move against each other. A distribution is good at the grain of the surface, where individual notes fall. What has been missing is not a better version of any one of them, but a single object that each can write to and read from. We make the parameter configuration that object. Each grammar symbol is mapped not to a note or a motif but to a complete distributional regime, meaning a distribution type for every musical parameter together with a canon ratio and a duration. A symbol therefore does not say play this. It says become this kind of music. We call this distribution-switching, and it is what allows a symbol to select a distribution that a ratio then time-scales.

Making that idea play on a real instrument requires one more layer. Once the pipeline has produced events, they must survive the mechanism, so a hardware abstraction layer treats the Disklavier's velocity-dependent latency and key-reset time as constraints on generation rather than as noise to be tolerated afterwards. The architecture that results runs from symbol to key in four layers, and the same convergence points that Nancarrow used as formal pillars are then wired back into it, so that when voices meet the system can switch distributions, and the music's own temporal structure decides where its texture changes.

We make four contributions.

\begin{enumerate}
    \item We introduce a hierarchical distribution-switching architecture that maps grammar symbols to distinct distributional regimes, and we show that compositional intent survives the pipeline by measuring the distortion each layer introduces and by ablating the L-system, the tempo canon, and the hardware compensation (Section~\ref{sec:results_ds}).
    \item We formalise a hardware abstraction layer that folds velocity-dependent latency and the key-reset time into the generation logic, so that every emitted trigger satisfies the per-key reset constraint, and we show that it remains effective even when the latency model is miscalibrated (Section~\ref{sec:hardware}).
    \item We locate, by direct measurement under a stated integration window rather than by assumption, the density of maximal separability between structured and random textures (near 25~notes/s) and the band over which single-domain melodic metrics lose that separability (40--100~notes/s), which delimits the regime where cross-domain constraints become the effective compositional lever (Section~\ref{sec:coherence}).
    \item We operationalise Convergence Point Calculus as a control interface between deterministic and stochastic structure, in which the tolerance $\epsilon$ becomes a compositional parameter that the composer sets (Section~\ref{sec:convergence}).
\end{enumerate}

The headline result concerns the mechanism's timing. Under the latency model of Section~\ref{sec:hardware}, pre-compensation takes the mean onset error from roughly 18~ms to below the Disklavier's own 1~ms scanning resolution. That figure is algorithmic, since it compares the compensator against the model it inverts rather than against a recorded piano, and what it establishes is that the compensation logic is sound and that any residual error is the model's rather than the software's. Whether the model matches a given instrument is a calibration question, and Section~\ref{sec:hardware} shows the compensation keeps most of its benefit when it does not. The rest of the paper reports how far the same claim can be pushed at the other three interfaces.

The quantitative results reported here are computational, deriving from algorithmic generation and statistical analysis of MIDI event streams rather than from controlled listening experiments. The statistical metrics employed (Shannon entropy, Kolmogorov-Smirnov distance, and Wasserstein distance) serve exclusively as computational measures of distributional structure. Although these metrics are theoretically motivated by auditory scene analysis \citep{bregman1990auditory} and information-theoretic aesthetics \citep{birkhoff1933aesthetic, meyer1957meaning}, their relationship with listener perception remains an open empirical question addressed in the proposed psychoacoustic protocol (Section~\ref{sec:future_protocol}). Where we draw on the perceptual literature, it is to motivate design choices and generate hypotheses, not to claim perceptual validity. This follows established practice. Xenakis' stochastic methods \citep{xenakis1992formalized}, analyses of Nancarrow \citep{gann1995music, nemire2014convergence}, and L-system composition \citep{manousakis2006musical} were assessed through formal, score-based, or authorial demonstration rather than controlled listening experiments. This paper extends that tradition with statistical pipeline validation and identifies a density-dependent loss of melodic discriminability that motivates future psychoacoustic research.

\section{Background and Related Work}
\label{sec:background}

\subsection{Tempo Canons and Convergence Points}

Nancarrow's roughly fifty Studies for Player Piano are the most sustained body of tempo-canon composition in Western music \citep{gann1995music}. \citet{thomas2000nancarrow} analysed the temporal projections in these canons, and the taxonomy she developed sorts them by where their convergence points fall, giving converging, diverging, arch, and diverging-converging types \citep{thomas1996temporal}. \citet{nemire2014convergence} worked that taxonomy into an explicit account of how convergence points are located and used. Rational tempo ratios converge periodically, as in the 3:4 canon of Study No.~15. Acceleration canons behave differently, and in Study No.~21 (``Canon X'') the two voices cross in tempo near the middle of the piece without that crossing being a convergence at all, the canonic convergence arriving later. Irrational ratios, such as the $\sqrt{2}$:2 of Study No.~33, admit no exact convergence. Nancarrow's own device for building tempo relations, the twelve-tempo scale of Study No.~37, follows Cowell's scales of rhythm \citep{cowell1996new}, in which tempo ratios are drawn from simple overtone ratios in the way pitch intervals are. \citet{callender2014performing} shows what irrational ratios cost in practice, since realising them requires rational approximation (7:5 for $\sqrt{2}$, for instance), the residual error accumulates with each repetition of the polyrhythm, and near-coincidences fall as little as 25~ms apart. That error analysis is what motivates the explicit tolerance parameter $\epsilon$ introduced in Section~\ref{sec:architecture}. All of this validation was formal and score-based, without listening experiments.

\subsection{Stochastic Composition}

Xenakis' \emph{Formalized Music} \citep{xenakis1992formalized} established the theoretical foundation for applying probability theory to musical composition. In \emph{Achorripsis}, he laid out a $7 \times 28$ matrix of 196 cells, seven timbral classes against twenty-eight units of time, and populated it by drawing event counts from a Poisson law. The ST program is a machine implementation of that scheme, and it derives the interval between successive attacks from an exponential law over a mean point density and the speed of a glissando from a Gaussian law, while pitch and duration follow their own distributions. The revised edition also covers Markov-chain state transitions (\emph{Analogique A/B}), dynamic stochastic synthesis of the waveform itself (the GENDY programs), and sieve theory for pitch-set generation. These approaches demonstrated that probability distributions could serve as first-class compositional objects, a principle that Amanous generalises through distribution-switching.

\subsection{L-Systems in Music}

Lindenmayer systems are rewriting grammars whose productions apply in parallel, replacing every symbol of a word at once. They were introduced to model cell division in multicellular organisms \citep{lindenmayer1968mathematicalI,lindenmayer1968mathematicalII}, and were later established as the standard formalism for plant morphology \citep{prusinkiewicz1990algorithmic}. \citet{manousakis2006musical} adapted L-systems for music composition, mapping alphabet symbols to musical parameters and production rules to transformations. \citet{worth2005growing} obtained hierarchical structure, in the Schenkerian sense of background, middleground, and foreground, from a stack-based musical interpretation of L-system strings. These approaches enable self-similar formal architectures but have not been integrated with tempo canon mathematics or electromechanical constraints.

\subsection{Algorithmic Composition Tools and Hardware-Aware Approaches}

Existing tools, including Max/MSP, SuperCollider, OpenMusic \citep{assayag1999computer}, and Essl's \emph{Lexikon-Sonate} \citep{essl2017algorithmic}, provide compositional primitives but lack a framework that simultaneously addresses symbolic structure, stochastic detail, and hardware constraints. Surveys of the field are organised one paradigm per chapter, with Markov models, grammars, chaotic systems, genetic algorithms, and neural networks treated separately \citep{nierhaus2009algorithmic}, which reflects how rarely a single system spans more than one of them. \citet{collins2018there} surveys new directions in large-scale algorithmic composition and gives particular attention to ``the mass scale of content generation now within reach''. Amanous pursues a concrete implementation by unifying three specific traditions within a single constrained pipeline.

Hardware-aware composition has been explored in specific artistic contexts. Ablinger's \emph{Quadraturen~III} renders speech and street noise through the piano's own sound repertoire, driven by frequency analysis of the recorded source, and no existing instrument could execute it. A Disklavier was tried and rejected, so an 88-solenoid \emph{Vorsetzer} was built for the piece \citep{ablinger2004quadraturen}. That instrument, the \emph{Autoklavierspieler}, is described by \citet{ritsch2011robotic}, who calibrates the force envelope and velocity curve of each key individually so that the piano can render intelligible speech. These projects demonstrate that hardware compensation is artistically productive but implement ad hoc solutions tied to specific works and specific instruments rather than a general framework.

A generative sequence model is the other plausible route to the compositional goals pursued here, and deep architectures such as Music Transformer \citep{huang2019music} and MuseNet \citep{payne2019musenet} produce piano music of considerable fluency \citep{briot2020deep}. Four properties of the present problem make a learned model the wrong instrument for it.

The first is the training distribution. A sequence model reproduces the statistics of the corpus it was trained on, and every large piano corpus consists of music written to be played by human hands. The densities, polyphony, and register spans targeted here lie outside that distribution by construction, so the corpus provides no support in exactly the region the system is meant to occupy. Learning what human pianists do is a poor way to leave the space of what human pianists can do.

The second is the hardware constraint. Velocity-dependent latency and key-reset time are hard physical bounds that must hold for every event in the output, not statistical tendencies to be approximated. A learned model can be biased towards satisfying such bounds through loss shaping or filtered sampling, but it cannot guarantee them, whereas an explicit constraint layer enforces them by construction and can be verified event by event (Section~\ref{sec:hardware}).

The third is interpretability of the control surface. The research questions posed here ask which layer of the pipeline distorts which parameter, and at what density a single-domain coherence metric loses its discriminative power. Answering them requires that the intended distribution be a stated object that the realised output can be compared against. In a learned model the intended distribution is implicit in the weights, so the layer-by-layer distortion analysis and the ablation study reported in Section~\ref{sec:results} would have no ground truth to measure against.

The fourth is the compositional relationship the system is designed to support. Amanous is conceived as an instrument that a composer parameterises rather than a model that a composer prompts (Section~\ref{sec:bg_creativity}). Reproducibility under a fixed seed, and the ability to attribute an audible change to a named parameter, are what make that relationship workable, and both are properties that explicit generative rules provide directly.

None of this argues that learned models are unsuited to composition in general. The claim is narrower. For a system whose purpose is to characterise the operational envelope of a physical instrument beyond the range of the available training data, an explicit and inspectable pipeline is the appropriate choice.

\subsection{Computational Creativity Frameworks}
\label{sec:bg_creativity}

\citet{boden2004creative} distinguished combinational, exploratory, and transformational creativity. \citet{wiggins2006preliminary} formalised that taxonomy as a search over conceptual spaces, defining a creative system in terms of a universe of possibilities together with rule sets that bound the space, traverse it, and evaluate what is found. \citet{jordanous2012standardised} operationalised creativity evaluation through fourteen components. \citet{loughran2017limitations} argued that creative systems should be assessed as processes, not simply as products. Amanous is conceptualised as an \emph{augmented compositional instrument} rather than an autonomous creative agent (Section~\ref{sec:creativity}). It does not seek autonomous creative agency, but extends the composer's reach into parameter spaces that exceed human physical capacity. The system's value lies in its predictive transparency. By providing feedback from pitch-class concentration and voice separation metrics after each generation (Section~\ref{sec:eval_framework}), it is intended to let the composer work with high-density textures against stated structural criteria, a claim the composer study proposed in Section~\ref{sec:future_protocol} would have to test.

\subsection{Disklavier Hardware Characteristics}
\label{sec:bg_hardware}

The hardware parameters the pipeline must respect are collected in Table~\ref{tab:hardware}. They come from three kinds of source, and the distinction matters. Velocity resolution, key scanning rate, key reset time, and key count are Disklavier Pro product specifications. Onset error is the one row backed by published measurement, and that measurement was made on an earlier instrument. The latency row is a modelling assumption of this study.

\begin{table}[htbp]
\tbl{Disklavier hardware parameters used by the pipeline, with the source of each. Product specifications describe the Disklavier Pro, and the measured row is from a Mark~II series grand.}
{\small\begin{tabular}{@{}llll@{}} \toprule
\textbf{Parameter} & \textbf{Value} & \textbf{Implication} & \textbf{Source} \\ \midrule
Velocity resolution & 1024 levels (10-bit) & XPMIDI protocol & Spec. \\
Key scanning rate & 800--1000~Hz & ${\sim}1$~ms resolution & Spec. \\
Key reset time & ${\sim}50$~ms & ${\sim}20$~Hz per key & Spec. \\
Key count & 88 & Max.\ simultaneous notes & Spec. \\
Onset error & $-20$ to $+30$~ms & Velocity-dependent & \citealp{goebl2003measurement} \\
Latency $L(v)$ & 30~ms (soft) to 10~ms (loud) & Modelled, Section~\ref{sec:hardware} & This study \\ \bottomrule
\end{tabular}}
\label{tab:hardware}
\end{table}

Velocity-dependent latency is the most consequential artefact, and it is also the least well characterised. Its direction follows from the mechanics of the action, since a key driven harder accelerates its hammer harder and the hammer reaches the string sooner. \citet{goebl2005piano} measured that relation on acoustic grands and fitted it with a power law in hammer velocity, reporting that a soft tone may take more than 200~ms longer to sound than a sudden fortissimo attack. Those magnitudes describe a key played by hand, so only the direction and the functional form carry over to a solenoid-driven action. Yamaha's own reproduction hardware is built around the same asymmetry. Its \emph{prelay} function buffers the incoming MIDI stream and fires the solenoids \emph{earlier for softer notes than for louder ones}, which is coherent only if a softly driven key is slower to speak \citep{goebl2003measurement}. The latency model of Section~\ref{sec:hardware} encodes that direction and that functional form.

What is missing is the curve itself. \citet{goebl2003measurement} report reproduction onset errors of $-20$ to $+30$~ms on a Disklavier grand, but their measurement compares a solenoid reproduction against the \emph{human} performance that was recorded, on an instrument whose internal compensation was left active, so it characterises a residual rather than the raw latency from MIDI command to hammer-string contact. No public calibration data set pairs MIDI velocity commands with measured acoustic-onset latency for any Disklavier model (Appendix~\ref{app:audit}). The latency model is therefore stated as a parameterised assumption to be calibrated per instrument rather than as an established curve. Two consequences follow, and Section~\ref{sec:hardware} takes both seriously. The compensation must degrade gracefully when the assumed curve is wrong, and on an instrument whose internal compensation cannot be defeated the residual curve, not the raw one, is what an external layer should apply.

\subsection{Auditory Scene Analysis and Density Thresholds}
\label{sec:bg_asa}

\citet{bregman1990auditory} established that listeners parse complex auditory scenes into coherent streams. \citet{huron2001tone} derived rules of voice-leading from those perceptual constraints, and his pitch proximity principle holds that a stream coheres when successive tones within it stay close in pitch, within about one to two semitones of each other in the region of greatest musical activity. That principle governs the integration of a single line rather than the separation of concurrent ones, and the perceptual literature does not supply a minimum registral gap that guarantees two simultaneous voices will segregate. The registral separation the pipeline enforces is therefore an engineering heuristic of this study, not a threshold taken from Huron.

At high event rates individual note identity dissolves, and \citet{roads2001microsound} places events between a few hundred microseconds and roughly 100~ms on the ``micro'' time scale, below which a fast sequence of pitched tones is heard as continuous texture rather than as rhythm. \citet{mcdermott2013summary} showed that the auditory system represents sound textures, such as rain, fire, and insect swarms, through time-averaged summary statistics rather than temporal fine structure. Whether the same holds for dense streams of pitched piano notes has not been tested. The density threshold for the melodic-to-textural transition has not been established by controlled experiment for piano timbres. Roads locates the fusion of discrete events into continuous texture at around 20~events/s \citep{roads2001microsound}, and the density sweep in Section~\ref{sec:coherence} runs from 10 to 200~notes/s so as to bracket that rate on both sides.

Information-theoretic measures of musical structure have a long history. \citet{birkhoff1933aesthetic} proposed that aesthetic value is the ratio of order to complexity, $M = O/C$. \citet{shannon1948mathematical} then provided the formalism on which the modern measures rest, and \citet{meyer1956emotion} argued that musical meaning arises from expectation and its inhibition, a reading he connected explicitly to information theory the following year \citep{meyer1957meaning}. Our framework adopts Levenshtein-based Melodic Coherence, entropy-based Pitch-Class Concentration, Kolmogorov-Smirnov-based Rhythmic Coherence, and the Wasserstein-based voice separation score as primary evaluation metrics (Section~\ref{sec:eval_framework}), maintaining explicit separation between computational measurement and perceptual inference.

\subsection{Towards a Common Parameter Space}
\label{sec:taxonomy}

Table~\ref{tab:taxonomy} maps the three compositional traditions to a common parameter space. In the table, $r_i$ is the tempo ratio of voice $i$ and $\tau_{\text{base}}$ the base IOI (Equation~\ref{eq:canon}). Each cell can be parameterised by a distribution type $\mathcal{D}$ with associated parameters, and the L-system symbol $s$ selects \emph{which} distribution governs each parameter, the distribution-switching mechanism formalised in Section~\ref{sec:architecture}.

\begin{table}[htbp]
\tbl{Common parameter taxonomy mapping Nancarrow, Xenakis, and L-system traditions to a shared distributional framework.}
{\resizebox{\linewidth}{!}{%
\begin{tabular}{@{}lllll@{}} \toprule
\textbf{Domain} & \textbf{Parameter} & \textbf{Nancarrow} & \textbf{Xenakis} & \textbf{Integrative (This Work)} \\ \midrule
\multirow{3}{*}{\textit{Temporal}} & Macro-form & Canon structure & Screen duration & L-system $\rightarrow$ symbol sequence \\
 & Tempo & Ratio (3:4, $e:\pi$) & n/a & $r_i$ per voice \\
 & IOI & $\tau_{\text{base}}/r_i$ & Exponential($\lambda$) & $\mathcal{D}_{\text{IOI}}(s)$ \\ \midrule
\multirow{2}{*}{\textit{Pitch}} & Register & Voice assignment & Gaussian($\mu, \sigma$) & $\mathcal{D}_{\text{pitch}}(s)$ \\
 & Pitch set & Cantus firmus & Uniform/weighted & Symbol-specific set \\ \midrule
\textit{Dynamic} & Velocity & Performance & Uniform/Gaussian & $\mathcal{D}_{\text{vel}}(s)$ \\ \midrule
\multirow{2}{*}{\textit{Structural}} & Convergence & Convergence-point calculus & n/a & $\epsilon$-triggered events \\
 & Self-similarity & n/a & n/a & Grammar depth $n$ \\ \bottomrule
\end{tabular}%
}}
\label{tab:taxonomy}
\end{table}

\section{Methods}
\label{sec:methods}

\subsection{System Architecture}
\label{sec:architecture}

The framework operates in four hierarchical layers (Figure~\ref{fig:architecture}), abbreviated throughout as L1 to L4. Layer~1 (L1) is the symbolic layer, where an L-system expands an axiom into a symbol sequence that fixes the macro-form. Layer~2 (L2) is the parametric layer, where each symbol is mapped to a complete parameter configuration, meaning a distribution type for each musical parameter together with a tempo-canon ratio set and a section duration. This mapping is the distribution-switching mechanism itself. Layer~3 (L3) is the numeric layer, where those distributions are sampled and the tempo canon is applied, yielding a list of time-stamped events. Layer~4 (L4) is the physical layer, where the hardware abstraction layer (HAL) pre-compensates velocity-dependent latency and enforces the electromechanical constraints of the instrument before MIDI is emitted.

The layer boundaries are also measurement points. Because each layer transforms the event stream, the distribution actually realised at the output can drift from the distribution the composer specified at L2. Sections~\ref{sec:results_ds} and \ref{sec:layer_analysis} quantify that drift by comparing the intended distribution against the realised one after L2, after L3, and after L4, so that each stage of the pipeline can be held separately accountable for the distortion it introduces.

\begin{figure}[htbp]
\centering
\resizebox{\linewidth}{!}{%
\begin{tikzpicture}[
    scale=0.93,
    transform shape,
    layer/.style={rectangle, draw=black, thick, minimum width=10cm, minimum height=1.2cm, align=center, font=\small},
    arrow/.style={-{Stealth[length=3mm]}, thick},
    lbl/.style={font=\footnotesize\itshape, text=black!65}
]

\node[ellipse, draw=black, thick, fill=yellow!10, minimum width=2.5cm, align=center] (Composer) at (-7, 5.75) {\textbf{Composer} \\ \footnotesize (Parameter Tuning)};

\node[layer, fill=blue!8] (L1) at (0,4.6) {\textbf{Layer 1: L-System Expansion} \\ $\omega \xrightarrow{P^n} S = s_1 s_2 \ldots s_m$};
\node[layer, fill=green!8] (L2) at (0,2.8) {\textbf{Layer 2: Parameter Mapping} \\ $s_i \xrightarrow{\mathcal{M}} \Theta_{s,g} = \langle \mathcal{D}_{\text{IOI}}, \mathcal{D}_{\text{pitch}}, \mathcal{D}_{\text{vel}}, \mathbf{r}^s, T^s \rangle$};

\draw[arrow, shorten <=2pt, shorten >=2pt]
  (Composer.east) .. controls +(1.2,0.2) and +(-1.2,0.2) .. (L1.west);
\draw[arrow, bend right=15] (Composer.south east) to (L2.west);

\node[layer, fill=orange!8] (L3) at (0,1.0) {\textbf{Layer 3: Event Generation} \\ Tempo canon $\otimes$ Stochastic sampling $\rightarrow \{(t_k, p_k, v_k)\}$};

\draw[arrow, dashed, gray, line width=0.8pt] (L3.west) -| (Composer.south) 
    node[near start, above, sloped, font=\scriptsize] {Coherence metrics};
\node[layer, fill=red!8] (L4) at (0,-0.8) {\textbf{Layer 4: Hardware Abstraction (HAL)} \\ Constraint enforcement $\otimes$ latency mapping $t \mapsto t - L(v)$};

\draw[arrow] (L1.south) -- (L2.north) node[midway, right, lbl] {symbol sequence};
\draw[arrow] (L2.south) -- (L3.north) node[midway, right, lbl] {distributions + ratios};
\draw[arrow] (L3.south) -- (L4.north) node[midway, right, lbl] {raw events};
\draw[arrow] (L4.south) -- ++(0,-0.7) node[below, font=\small\bfseries] {MIDI Output to Disklavier};

\node[font=\footnotesize, anchor=east] at (-7.9,4.6) {Symbolic};
\node[font=\footnotesize, anchor=east] at (-7.9,2.8) {Parametric};
\node[font=\footnotesize, anchor=east] at (-7.9,1.0) {Numeric};
\node[font=\footnotesize, anchor=east] at (-7.9,-0.8) {Physical};

\draw[arrow, dashed, blue!60, line width=1.1pt] (L3.east) -- ++(1.5,0) |- (L2.east) 
    node[midway, right, lbl, text width=2.5cm, blue!80] {\textbf{Feedback Loop:}\\CP events trigger\\dist-switching};
\end{tikzpicture}%
}
\caption{The four-layer architecture. The blue dashed path is the feedback loop by which convergence points detected at Layer~3 trigger distribution switches at Layer~2, and the grey dashed path returns coherence metrics to the composer.}
\label{fig:architecture}
\end{figure}

\subsubsection{Formal Definitions}

\begin{definition}[Dynamic Distribution-Switching Mapping]
For a symbol $s$ at the depth of derivation $g$, the parameter configuration $\Theta_{s,g}$ is determined by a mapping function $\mathcal{M}: \Sigma \times \mathbb{N} \rightarrow \mathbb{P}$, where $\mathbb{P}$ is the space of parameter configurations:
\begin{equation}
    \Theta_{s,g} = \mathcal{M}(s, g) = \langle \mathcal{D}_{\text{IOI}}^{s}(g), \; \mathcal{D}_{\text{pitch}}^{s}(g), \; \mathcal{D}_{\text{vel}}^{s}, \; \mathbf{r}^s, \; T^s \rangle
\end{equation}
where $\mathcal{D}_*(g)$ scales parameters (e.g.\ $\lambda$ for IOI, $\sigma$ for pitch) as a function of depth $g$, ensuring that the hierarchical structural complexity of Layer~1 is preserved in the distributional grain of Layer~2. Where the derivation depth is not at issue we write $\Theta_s$ for $\Theta_{s,g}$.
\end{definition}

\begin{definition}[Convergence Point]
\label{def:cp}
For voices with onset timelines $T_i(n)$ and $T_j(m)$, a convergence point exists within tolerance $\epsilon$ when:
\begin{equation}
    \exists \, n, m \in \mathbb{N} : |T_i(n) - T_j(m)| < \epsilon
\end{equation}
\end{definition}

\subsubsection{Core Mechanism: Distribution-Switching}

Each L-system symbol selects a different distribution \emph{type} for each musical parameter, not just different parameter values within a fixed family. For example, symbol $A$ might map IOI to a constant (deterministic rhythm), while symbol $B$ maps IOI to an exponential distribution (stochastic rhythm). Mode-switching between generative regimes exists in modular environments (Max/MSP, SuperCollider) and is implicit in Xenakis' matrix-based sectional organisation. The contribution here is end-to-end integration, in that these switches are (a)~triggered deterministically by an L-system grammar, (b)~combined with tempo-canon time-scaling, and (c)~passed through a compensation layer before actuation.

\subsubsection{L-System Macro-Form Generation}

For canonical instantiation with alphabet $\Sigma = \{A, B\}$, axiom $\omega = A$, and production rules $P: A \rightarrow AB, \; B \rightarrow A$, the L-system generates the Fibonacci-growth sequence $A \rightarrow AB \rightarrow ABA \rightarrow ABAAB \rightarrow ABAABABA \rightarrow \ldots$\,. The L-system was selected for its structural isomorphism with tempo canons, since both rely on recursive scaling, so that macro-form and micro-form share a common hierarchical logic. This grammar-driven expansion generates a structural narrative in which recursion depth modulates the transition between sparse and dense regimes, producing hierarchical self-similarity that simple stochastic state-transition models cannot replicate.

Layers~1 and~2 are coupled through recursion-depth modulation. The expansion stage is implemented so that each output symbol is tagged with its \emph{generation} (recursion depth). Axiom symbols have generation~0, and symbols produced in the $g$-th rewrite have generation~$g$. Layer~2 then uses this tag to modulate the parameter distributions, so that a higher depth is assigned \emph{a denser} IOI (e.g.\ smaller scale for exponential $\mathcal{D}_{\text{IOI}}$, so higher effective $\lambda$) and \emph{a wider} pitch range (e.g.\ larger $\sigma$ for Gaussian $\mathcal{D}_{\text{pitch}}$ or broader support for uniform). Because an L-system rewrites every symbol in parallel, all symbols of a fully expanded string carry the generation of the final rewrite, so within one piece the tag scales every section by the same factor, and it differentiates pieces expanded to different depths rather than sections within a piece. A tag that distinguished sections by their position in the derivation tree is a natural extension and is not implemented here. The effect of the modulation as implemented is evaluated in Section~\ref{sec:ablation} via Lempel-Ziv complexity on the rendered MIDI, by comparing depth-weighted and symbol-only pipelines.

\subsubsection{Tempo Canon Implementation}

For a canon with $K$ voices in tempo ratios $r_1 : r_2 : \ldots : r_K$, the voice $i$ generates events at cumulative times:
\begin{equation}
    T_i(k) = T_0 + k \, \frac{\tau_{\text{base}}}{r_i}
\label{eq:canon}
\end{equation}
where $T_0$ is the section start time and $\tau_{\text{base}}$ is the base IOI. For acceleration/deceleration canons, $\text{IOI}_i(k) = \text{IOI}_i(0) \cdot \alpha_i^k$.

\subsubsection{Stochastic Event Generation}

Within each section, timing follows the distribution the symbol specifies, which may be a constant IOI $\tau_k = c$ (deterministic), an exponential IOI $\tau_k \sim \text{Exp}(\lambda)$ (a Poisson process), or an inhomogeneous Poisson process with time-varying rate $\lambda(t)$. Pitch draws from symbol-specific distributions (chromatic or scale-weighted). Velocity follows constant, uniform, or Gaussian distributions.

\subsubsection{Voice Allocation and Collision Avoidance}
To respect the electromechanical key reset time ($\sim$50~ms), Layer~4 applies a key-reset mask after latency pre-compensation. For each key, across all voices, triggers are taken in the order in which the solenoid would be driven, and any trigger that falls within 50~ms of the last retained trigger on the same key is suppressed. A suppressed event is dropped rather than merged or delayed, so the timing of every surviving event is untouched. The mask is always active, and its cost is reported with the results. It removes events, and what it removes is a measured quantity rather than a hidden one.

\subsubsection{Event Generation Algorithm}

Algorithm~\ref{alg:generation} presents the pipeline in simplified form. Recursion-depth modulation enters through the configuration $\Theta_s$ of each symbol occurrence, and the full Layer~4 stage is given in Algorithm~\ref{alg:latency} (Appendix~\ref{app:algorithm}).

\begin{algorithm}[htbp]
\caption{Hierarchical Distribution-Switching Event Generation}
\label{alg:generation}
\begin{algorithmic}[1]
\Require Alphabet $\Sigma$, axiom $\omega$, rules $P$, depth $n$, mappings $\{\Theta_s\}_{s \in \Sigma}$
\Ensure Time-stamped event list $E = \{(t, p, v, d)\}$
\State $S \gets \textsc{LSystemExpand}(\omega, P, n)$
\State $E \gets \emptyset$;\; $t_{\text{cur}} \gets 0$
\For{each symbol $s$ in $S$}
    \State $(\mathcal{D}_{\text{IOI}}, \mathcal{D}_p, \mathcal{D}_v, \mathbf{r}, T) \gets \Theta_s$
    \For{each voice $i = 1, \ldots, |\mathbf{r}|$}
        \State $t_i \gets t_{\text{cur}}$
        \While{$t_i < t_{\text{cur}} + T$}
            \State $\tau \gets \textsc{Sample}(\mathcal{D}_{\text{IOI}}) \;/\; r_i$ \Comment{Tempo-scaled IOI}
            \State $p \gets \textsc{Sample}(\mathcal{D}_p)$;\quad $v \gets \textsc{Sample}(\mathcal{D}_v)$
            \State $v \gets \textsc{Clamp}(v, 0, 1023)$
            \State $t_{\text{adj}} \gets t_i - L(v)/1000$ \Comment{See Section~\ref{sec:hardware}}
            \State $E \gets E \cup \{(t_{\text{adj}}, p, v, \tau)\}$
            \State $t_i \gets t_i + \tau$
        \EndWhile
    \EndFor
    \State $t_{\text{cur}} \gets t_{\text{cur}} + T$
\EndFor
\State \Return \textsc{SortByOnset}($E$)
\end{algorithmic}
\end{algorithm}

\subsubsection{Worked Example: Configuring a Composition}
\label{sec:worked_example}

The parameters a composer actually sets are best seen on a concrete piece. The canonical instantiation used throughout Section~\ref{sec:results} is configured as follows, and it is the same configuration reported in Table~\ref{tab:symbol_config}.

The composer first fixes the macro-form at Layer~1 by choosing an alphabet, an axiom, production rules, and a recursion depth. Here the alphabet is $\Sigma = \{A, B\}$, the axiom is $\omega = A$, the rules are $A \rightarrow AB$ and $B \rightarrow A$, and the depth is 4. Expansion yields the section sequence $ABAABABA$, so the piece will have eight sections, five of type $A$ and three of type $B$, in that order. Nothing about the sound has been decided at this point. The grammar has fixed only the order and proportion of the sections, which is the composer's control over large-scale form.

The composer then decides, at Layer~2, what each symbol means. This is where distribution-switching happens, and it is the substantive compositional act. Symbol $A$ is defined as a deterministic regime, taking a constant inter-onset interval, a pitch set restricted to the seven pitch classes of C major, a constant velocity of 640, a tempo-canon ratio of 3:4, and a section duration of about 10~seconds. Symbol $B$ is defined as a textural regime, taking an exponentially distributed inter-onset interval, the full chromatic pitch set, a velocity drawn uniformly from $[400, 900]$, a canon ratio of $1{:}\sqrt{2}$, and a section duration of 8~seconds. The two symbols therefore differ in the \emph{type} of distribution assigned to each parameter, not merely in the parameter values, which is what makes the resulting sections statistically separable rather than simply louder or faster. The densities that follow from these settings, once Layer~2's depth modulation is applied, are 64.3 and 117.4~notes/s respectively, so the piece alternates between two regimes that both lie beyond a pianist's reach and that remain distinguishable from one another.

Each symbol also carries the recursion depth at which it was produced. In this piece every section is produced by the fourth rewrite, so all eight carry depth 4 and receive the same scaling, a denser inter-onset distribution and a wider pitch range than the base configuration specifies. The composer sets the depth-scaling rule once, and a piece expanded to a different depth is scaled accordingly without any section being respecified by hand.

Layers~3 and 4 require no further compositional input. Layer~3 samples the chosen distributions and applies the canon ratios, producing 6{,}591 time-stamped events across 74~seconds. Layer~4 pre-compensates each onset by the latency its velocity implies and then enforces the key-reset constraint, which suppresses 560 of those events and leaves 6{,}031. What the composer tuned amounts to a grammar, a recursion depth, and one parameter block per symbol. Everything audible follows from those choices, and a fixed random seed makes the realisation reproducible.

\subsection{Evaluation Framework}
\label{sec:eval_framework}

\subsubsection{Evaluation Metrics}

The following metrics serve as diagnostic tools to monitor the transmission of structural intent. They do not measure absolute performance but quantify the relative parameter dependence within different density regimes.

\emph{Melodic Coherence} (MC) is one minus the normalised Levenshtein edit distance between two pitch contours, each encoded as an Up/Down/Same sequence:
\begin{equation}
    \text{MC}(X, Y) = 1 - \frac{d_{\text{Lev}}(\text{contour}(X), \text{contour}(Y))}{\max(|\text{contour}(X)|, |\text{contour}(Y)|)}
\label{eq:mc}
\end{equation}

\emph{Single-Voice Coherence} (SVC) characterises one voice in isolation. Melodic Coherence compares two sequences, so for a single voice we use the normalised entropy of its pitch-interval distribution. For a pitch sequence $X = (p_1, \ldots, p_n)$, let $\iota_k = \mathrm{clip}(p_{k+1} - p_k, -12, +12)$ for $k = 1, \ldots, n-1$ be its melodic intervals in semitones, truncated to $[-12, +12]$ and so taking one of $B = 25$ interval classes. Writing $\hat{q}(b)$ for the empirical frequency of class $b$,
\begin{equation}
    \text{SVC}(X) = 1 - \frac{H(\hat{q})}{\log_2 25}, \qquad H(\hat{q}) = -\sum_{b} \hat{q}(b) \log_2 \hat{q}(b)
\label{eq:svc}
\end{equation}
SVC equals 1 when every interval is identical and falls towards 0 as the interval distribution approaches uniformity over the 25 classes. It is a single-sequence quantity and is not the pairwise MC of Equation~\ref{eq:mc}. MC is used for the same-symbol versus cross-symbol pair comparisons of Section~\ref{sec:results_ds} and for the ablation of Section~\ref{sec:ablation}. SVC is computed over a fixed event count and so cannot vary with density, which is why the density sweep of Section~\ref{sec:coherence} measures transition retention instead.

\emph{Rhythmic Coherence} (RC) is one minus the Kolmogorov-Smirnov (KS) distance between two IOI distributions:
\begin{equation}
    \text{RC}(X, Y) = 1 - D_{KS}(F_{\text{IOI}}^X, F_{\text{IOI}}^Y)
\end{equation}

\emph{Pitch-Class Concentration} (PCC) is an entropy-based measure of how concentrated the pitch-class distribution is:
\begin{equation}
    \text{PCC} = 1 - \frac{H(\text{pitch-class})}{\log_2 12}
\label{eq:pcc}
\end{equation}
Higher PCC indicates a stronger statistical bias towards specific pitch classes. PCC is used here as a measure of distributional entropy.

The \emph{Voice Separation Score} (VSS) is the mean Wasserstein distance between two voices in pitch, velocity, and log-IOI:
\begin{equation}
    \text{VSS} = \frac{1}{3}\bigl(W_1(F_p^i, F_p^j) + W_1(F_v^i, F_v^j) + W_1(F_\tau^i, F_\tau^j)\bigr)
\end{equation}

\emph{Weighted VSS} (wVSS) is a component-weighted variant in which the weights $w_*$ act as structural importance factors and $W_{\text{pitch}}$, $W_{\text{vel}}$, and $W_{\text{temporal}}$ are the three Wasserstein components of VSS. The weights are derived from a reference high-density baseline to provide a consistent benchmark for cross-domain comparison, avoiding sequence-specific bias:
\begin{equation}
    \text{wVSS} = w_{\text{pitch}} \cdot W_{\text{pitch}} + w_{\text{vel}} \cdot W_{\text{vel}} + w_{\text{temporal}} \cdot W_{\text{temporal}}
\label{eq:wvss}
\end{equation}
Because the weights are derived from the same experimental data to which wVSS is subsequently applied (Section~\ref{sec:coherence}), this metric characterises the relative contribution of each domain within the tested condition. Split-half cross-validation of the weights is reported in Section~\ref{sec:wvss_validation}.

To control for scale differences across parameter domains, we define a range-normalised variant, nwVSS, that divides each Wasserstein component by the theoretical range of its domain before weight computation.
\begin{equation}
    \text{nwVSS} = w_{\text{pitch}} \cdot \frac{W_{\text{pitch}}}{R_{\text{pitch}}} + w_{\text{vel}} \cdot \frac{W_{\text{vel}}}{R_{\text{vel}}} + w_{\text{temporal}} \cdot \frac{W_{\text{temporal}}}{R_{\text{temporal}}}
\label{eq:nwvss}
\end{equation}
where $R_{\text{pitch}} = 127$ (MIDI note range), $R_{\text{vel}} = 1023$ (XPMIDI range), and $R_{\text{temporal}}$ is the theoretical or observed range of the log-IOI domain (e.g.\ $\ln(10/0.001) \approx 9.21$ for IOI in seconds). An identical weight-estimation procedure is applied to the normalised components to obtain nwVSS weights, which allows a check against scale-driven dominance.

\subsubsection{Statistical Analysis}

All comparisons used two-tailed tests with $\alpha = .05$. Effect sizes are reported as Cohen's $d$ with 95\% confidence intervals (CI) for $t$-tests, and rank-biserial $r$ for Mann-Whitney $U$ tests, signed by the direction of change from the first condition to the second. The confidence intervals for $d$ were calculated by inversion of the noncentral $t$ distribution. Table~\ref{tab:stats_full} in Appendix~\ref{app:stats} collects every test in one place.

\subsubsection{Scope and Validation Philosophy}

The statistical tests assess \emph{pipeline fidelity}, that is, whether the four-layer processing chain faithfully preserves intended distributional separations, thereby transmitting design intent without degradation. Layer-by-layer degradation analysis (Section~\ref{sec:layer_analysis}) extends this by quantifying the distortion each processing stage introduces and identifying the most vulnerable parameters.

\subsubsection{Experimental Conditions}

Six conditions were evaluated. Each isolates one link in the chain that runs from grammar symbol to actuation-ready MIDI, and together they cover the four contributions stated in Section~\ref{sec:intro}. The configuration of each is given below so that the results in Section~\ref{sec:results} can be read against the settings that produced them.

The \emph{pipeline fidelity} condition (Section~\ref{sec:results_ds}) is the canonical instantiation described in the worked example above. The grammar $A \rightarrow AB$, $B \rightarrow A$ expands the axiom $A$ to depth 4, giving the section sequence $ABAABABA$, with two voices per section and the symbol configurations of Table~\ref{tab:symbol_config}. The run lasts 74~seconds and yields $N = 6{,}031$ events after Layer~4. This condition supplies both the end-to-end separation tests and the layer-by-layer degradation analysis, since the intended distribution for every parameter is known by construction and can be compared against the realised one after each layer.

The \emph{beyond-human demonstration} (Section~\ref{sec:results_ds}) tests whether specifications that no pianist can execute survive the pipeline as MIDI. It comprises three sections, each targeting a different physical limit, namely 40-note simultaneities at 500~ms spacing, a 30~Hz trill realised through multi-key alternation because single-key repetition is bounded by the key reset time, and a six-octave arpeggio at a 25~ms inter-onset interval. Here the quantity of interest is specification error rather than a distributional comparison.

The \emph{hardware compensation} condition (Section~\ref{sec:hardware}) evaluates the latency model of Layer~4 over $N = 526$ notes spanning the usable velocity range, comparing the onset timing achieved with pre-compensation against the timing that the same events would receive without it. Velocities are drawn across the full range of Yamaha's 10-bit extended-precision MIDI (XPMIDI) so that the velocity-dependent portion of the latency curve is exercised rather than only its centre.

The \emph{density sweep} (Section~\ref{sec:coherence}) locates the density of maximal structured-versus-random separability. Fourteen density levels are generated, at 10, 15, 20, 25, 28, 30, 40, 50, 60, 80, 100, 120, 150, and 200~notes/s, and at each level a structured stream and a random null stream of 200 events are generated in each of 40 trials and compared under a fixed temporal integration window. The levels are spaced more finely between 20 and 30~notes/s because the fusion of discrete events into texture is reported at around 20~events/s \citep{roads2001microsound}, which is where a break point was expected.

The \emph{cross-domain constraints} condition (Section~\ref{sec:coherence}) asks which parameter domain does the work of keeping voices separable once density is high. Four voices generate 500 events per condition. The baseline applies no separation constraint. A pitch constraint then enforces a registral separation greater than 5~semitones between voices. A combined constraint additionally stratifies the voices into disjoint velocity bands and distinct temporal profiles. A final variant imposes an explicit pitch-velocity coupling, $v = 200 + 12.5 \times (p - 40)$ over MIDI notes 40 to 105, which makes dynamics a deterministic function of register.

The \emph{convergence point} condition (Section~\ref{sec:convergence}) tests the feedback path from Layer~3 back to Layer~2. The discrete case uses a 3:4 rational canon, with two canon voices at inter-onset intervals of 1.000~s and 0.750~s plus one Poisson voice, running 60~seconds with the distribution switch armed at the convergence at $t = 30$~s. The continuous case uses an $e{:}\pi$ canon, an irrational ratio for which no exact convergence exists, running 30~seconds for 761 events, so that convergence must be detected through the tolerance parameter $\epsilon$ rather than by exact coincidence.

All conditions used deterministic random seeds, with seed 42 for the canonical instantiation, so that any reported figure can be regenerated exactly. The system was implemented in Python with microsecond-resolution timestamps. Supplementary audio (Excerpts~1--4) accompanies this article as a compressed archive. Source code is at \url{https://github.com/joonhyungbae/amanous}.

\section{Results}
\label{sec:results}

\subsection{Distribution-Switching Validation}
\label{sec:results_ds}

\subsubsection{Canonical Instantiation}

Grammar $A \rightarrow AB$, $B \rightarrow A$ (axiom $A$, depth~4) produced the string $ABAABABA$. Symbol configurations are given in Table~\ref{tab:symbol_config}. Under seed 42, Layer~3 produces 6{,}591 time-stamped events over 74.03~seconds. The key-reset mask of Layer~4 suppresses 560 of them (8.5\%), and the $N = 6{,}031$ that remain are the output analysed below. The realised densities exceed the nominal ones because Layer~2 modulates each symbol's parameters by its recursion depth, which at depth~4 makes every section denser than the base configuration specifies. With depth modulation disabled the same grammar yields 3{,}555 events at Layer~3 and 3{,}383 after the mask, and that variant is the symbol-only control used in the ablation of Section~\ref{sec:ablation}.

The distribution-switching signature is directly visible in the output. Symbol $A$ realises exactly the seven pitch classes of C major and a single velocity value, while symbol $B$ realises all twelve pitch classes and 63 distinct velocities. The symbols select distribution \emph{types}, not parameter values within one type, and that difference survives to the event stream.

\begin{table}[htbp]
\tbl{Symbol-to-parameter mapping for the canonical instantiation. Each row is one component of the configuration $\Theta_s$ that Layer~2 assigns to a symbol.}
{\begin{tabular}{@{}lll@{}} \toprule
\textbf{Parameter} & \textbf{Symbol $A$ (Deterministic)} & \textbf{Symbol $B$ (Textural)} \\ \midrule
Tempo canon ratio & 3:4 & $1{:}\sqrt{2}$ \\
IOI distribution & Constant & Exponential \\
Pitch set & C major (7 PCs) & Chromatic (12 PCs) \\
Velocity & Constant ($v = 640$) & Uniform $[400, 900]$ \\
Section duration ($T^s$) & 10~s & 8~s \\ \midrule
Realised density & 64.3 notes/s & 117.4 notes/s \\ \bottomrule
\end{tabular}}
\label{tab:symbol_config}
\end{table}

\subsubsection{Density Bifurcation}

Deterministic sections ($A$) averaged 64.3~notes/s aggregate, and textural sections ($B$) averaged 117.4~notes/s. The 53.1~notes/s gap separates the two regimes without overlap across sections (exact Mann-Whitney $U = 0.0$, $p = .036$, $n = 5 + 3$ sections). Both regimes lie beyond human reach, which is the point of the system, and they nevertheless remain statistically distinct from one another. The pitch streams in the deterministic sections were separated by 12.4~semitones, comfortably above the 5-semitone registral separation the pipeline enforces as a design heuristic (Section~\ref{sec:bg_asa}).

\subsubsection{End-to-End Fidelity: Coherence Metrics}

\begin{table}[htbp]
\tbl{Pipeline fidelity: same-symbol vs.\ cross-symbol coherence ($n_{\text{same}} = 13$, $n_{\text{cross}} = 15$ pairs).}
{\small\begin{tabular}{@{}lccccl@{}} \toprule
\textbf{Metric} & \textbf{Same} & \textbf{Cross} & \textbf{Gap} & \textbf{Test} & \textbf{Effect [95\% CI]} \\ \midrule
Melodic (MC) & $.770 \pm .007$ & $.673 \pm .012$ & .097 & $t(26)=25.53$*** & $d=9.67$ $[7.03, 12.31]$ \\
Rhythmic (RC) & $.820 \pm .057$ & $.476 \pm .007$ & .344 & $t(26)=23.07$*** & $d=8.74$ $[6.33, 11.15]$ \\ \midrule
\multicolumn{6}{@{}l}{\textit{Pitch-Class Concentration by symbol ($n = 5$ A-sections, $3$ B-sections)}} \\
PCC & $.230 \pm .003$ & $.004 \pm .003$ & .225 & $U=15.0$* & $d=75.98$ \\ \bottomrule
\end{tabular}}
\tabnote{***$p < .001$; *$p < .05$. The PCC row compares symbols rather than section pairs, so its columns read as symbol $A$ and symbol $B$.}
\label{tab:coherence}
\end{table}

Table~\ref{tab:coherence} reports the end-to-end comparisons. Same-symbol section pairs are more coherent than cross-symbol pairs on both the melodic and the rhythmic metric. Low-entropy pitch selection (C major, $n = 5$ A-sections) was also compared against high-entropy selection (chromatic, $n = 3$ B-sections). PCC was $.230 \pm .003$ for the C-major sections and $.004 \pm .003$ for the chromatic ones (exact Mann-Whitney $U = 15.0$, $p = .036$, rank-biserial $r = -1.00$, $d = 75.98$). The direction is the expected one, since restricting the pitch set to seven classes concentrates the pitch-class distribution while the chromatic set spreads it towards uniformity. With eight sections the test has little power to do more than separate the two groups, which it does without overlap.

The 28 pairs are formed from the same eight sections and are therefore not independent, so the degrees of freedom and the interval on $d$ should be read descriptively rather than inferentially. The effect sizes are large because the contrasts are ones the configuration specifies, and what they measure is how cleanly a designed separation survives the pipeline.

The velocity contrast is qualitative rather than statistical. Symbol $A$ emits a single velocity by construction, so its distribution has zero spread and no effect size is defined against symbol $B$'s uniform draw.

\subsubsection{Pipeline Degradation: Per-Layer Distortion}
\label{sec:layer_analysis}

\begin{table}[htbp]
\tbl{Layer-by-layer distributional degradation, as KS distance between intended and realised distributions (6{,}591 events up to the compensation stage, 6{,}031 after the key-reset mask).}
{\small\begin{tabular}{@{}lccccl@{}} \toprule
 & & & \multicolumn{2}{c}{\textbf{After L4}} & \\ \cmidrule(lr){4-5}
\textbf{Parameter} & \textbf{After L2} & \textbf{After L3} & \textbf{comp.} & \textbf{+ mask} & \textbf{Stage of largest increase} \\ \midrule
IOI (Symbol $A$) & 0.000 & 0.000 & 0.000 & 0.087 & L4 (key-reset mask) \\
IOI (Symbol $B$) & 0.036 & 0.032 & 0.322 & 0.291 & L4 (latency compensation) \\
Pitch (Symbol $A$) & 0.039 & 0.039 & 0.039 & 0.040 & None (fixed at L2) \\
Pitch (Symbol $B$) & 0.035 & 0.035 & 0.035 & 0.031 & None (fixed at L2) \\
Velocity (Symbol $A$) & 0.000 & 0.000 & 0.000 & 0.000 & None (constant) \\
Velocity (Symbol $B$) & 0.045 & 0.045 & 0.045 & 0.045 & None (fixed at L2) \\ \bottomrule
\end{tabular}}
\tabnote{KS distances below 0.05 indicate negligible distributional distortion. ``comp.'' is latency pre-compensation alone, and ``+ mask'' is the complete Layer~4.}
\label{tab:degradation}
\end{table}

To go beyond confirming that the designed separations survive the pipeline, we quantified \emph{how much} distributional distortion each processing stage introduces and \emph{which parameters} are the most vulnerable.

For each of the three primary musical parameters (IOI, pitch, velocity), we measured the Kolmogorov-Smirnov distance between the intended distribution and the actual output at three measurement points, namely after Layer~2 (a direct draw from the configured distributions, before the canon is applied), after Layer~3 (tempo canon and event generation applied), and after Layer~4, which is reported in two stages, first with latency pre-compensation alone and then with the key-reset mask added. Table~\ref{tab:degradation} reports the results. Each column is a cumulative KS distance, so the distortion a layer contributes on its own is the increase over the preceding column, and the final column names the layer with the largest such increase.

The distortion is concentrated at one interface, and it is not the one we expected. Tempo scaling at Layer~3 introduces essentially none, because it divides every interval of a voice by the same ratio, which maps the intended distribution onto its scaled counterpart exactly. The stage that distorts is Layer~4, and its two stages act differently. Pre-compensation shifts each event earlier by an amount that depends on its velocity, so a section with a constant velocity is shifted rigidly and its intervals are untouched, while a section drawing velocity from a uniform distribution has every interval perturbed by the difference between two independent velocity draws. Symbol $A$, whose velocity is constant, accordingly shows a KS distance of exactly zero through the compensation stage, and symbol $B$ rises from $0.032$ to $0.322$. The key-reset mask then acts on both symbols, because removing an event merges the two intervals on either side of it. It raises the IOI distance of symbol $A$ from zero to $0.087$, and it leaves that of symbol $B$ close to where compensation put it ($0.291$). Pitch and velocity are fixed at Layer~2 and no value is rewritten downstream, so their small deviations from the intended distribution ($D_{KS} \leq 0.045$, attributable to finite sampling and to the pitch-set snapping the design specifies) are constant up to the mask and move by less than $0.005$ across it. This has a direct consequence for the design of convergence points (Section~\ref{sec:convergence}). A switch that changes the pitch set or the velocity distribution is transmitted to the instrument essentially untouched, whereas a switch that changes the IOI distribution is the one whose realised form depends on the velocities the section happens to draw. The composer who wants a predictable contrast at a convergence point should reach for pitch or velocity, and the composer who wants the hardware to leave its own trace should reach for IOI.

\subsubsection{End-to-End Demonstration}

Pipeline fidelity is one thing, and reaching the specifications that motivated the system is another. Table~\ref{tab:demo} reports three sections written against the physical limits named in Section~\ref{sec:intro}, namely a forty-note simultaneity, a trill above the rate a single key can be re-struck, and an arpeggio spanning six octaves at an inter-onset interval no hand could sustain. Each was specified first and generated second, and the MIDI stream realises each specification exactly. That exactness is a property of the generator rather than a result, since the events are constructed from the specification. What the table does show is that the three sections remain distributionally distinct from one another once generated, and that each specification passes through Layer~4 intact, since the key-reset mask suppresses no event in any of the three sections. The trill is the instructive case. A single key cannot be re-struck at the target rate because of the reset time of the action, so the specification is written as an alternation across two adjacent keys, which brings the rate per key down to 15~Hz and the smallest same-key interval to 61~ms. Written for a single key, the same 30~Hz stream would lose every second note to the mask.

\begin{table}[htbp]
\tbl{Beyond-human-density specifications and achieved performance at the MIDI-event level ($N = 3$ sections).}
{\begin{tabular}{@{}lll@{}} \toprule
\textbf{Section} & \textbf{Specification} & \textbf{Realised in MIDI} \\ \midrule
Polyphony & 40-note chords at 500~ms & 40 notes, 500~ms \\
Repetition & 30~Hz trill\textsuperscript{a} & 30~Hz, alternating keys \\
Speed/Span & 6-octave arpeggio, 25~ms IOI & 72 semitones, 25~ms \\ \midrule
\multicolumn{2}{@{}l}{Between-section KS distances} & all $p < .001$ \\ \bottomrule
\end{tabular}}
\tabnote{\textsuperscript{a}Single-key repetition is limited to $\sim$20~Hz by the key reset time, so the 30~Hz target requires multi-key alternation. The key-reset mask suppresses no event in these sections. All sections are rendered as algorithmic MIDI output (Excerpt~1).}
\label{tab:demo}
\end{table}

\subsubsection{Musical Analysis of Generated Output}
\label{sec:musical_analysis}

To complement statistical validation, we briefly describe the structural character of the generated output, as documented in the supplementary materials (Excerpts~1--4). The excerpts are software renders of the generated MIDI through a sampled piano, not recordings of an instrument. This description is based on the distributional and parametric design.

The 74-second canonical instantiation (Excerpt~3) exhibits a clear alternation between two contrasting textural states. In the deterministic sections ($A$), two canon voices in a 3:4 tempo ratio produce interlocking rhythmic patterns separated by 12.4~semitones in register. The resulting texture is contrapuntal, with two distinct melodic streams, each with a regular rhythmic profile, that produce hocketing interactions in which the faster voice fills the temporal gaps of the slower one. The musical effect is comparable to the layered metrical processes in Nancarrow's Study No.~15 (Canon 3/4), where the rational ratio produces a predictable periodicity of alignment, but the individual voices are kept apart by design through consistent registral separation.

At each A$\rightarrow$B transition, the texture undergoes an abrupt qualitative shift. The onset of a textural section ($B$) replaces the deterministic two-voice counterpoint with a dense stochastic cloud. The exponential IOI distribution produces irregular, clustered onsets across all 12 pitch classes, with velocity varying over the $[400, 900]$ XPMIDI range. The intended textural effect (documented in Excerpt~3) is designed to resemble the granular textures in Xenakis' \emph{Pithoprakta}, where individual string glissandi fuse into a continuous sound mass. Unlike Xenakis' orchestral forces, however, the design relies on the piano's percussive attack envelope to maintain a degree of grain articulation even at the 117.4~notes/s of a $B$ section, with individual key strikes expected to remain partially audible as a shimmering, pointillistic texture rather than a fully fused continuum.

In the 34-second demonstration (Excerpt~1), the polyphony section (40-note chords) produces massive vertical sonorities whose harmonic density exceeds any pianistic chord voicing, an organ-like effect, with the full registral span of the instrument activated simultaneously. The 30~Hz multi-key trill section places its repetition rate in the region that Roads' micro-time-scale framework associates with fusion, between rapid figuration and a sustained timbre. The 6-octave arpeggio at 25~ms IOI traverses the keyboard faster than any human hand could sweep, producing a glissando-like wash of pitch that functions as a timbral gesture rather than a melodic one.

A separate composition, \emph{Phase Music, a Minimalist Study} (80~s; Excerpt~2), applies the same pipeline to a Reich-inspired phase shift design (pentatonic set, 1:1.01 tempo drift between voices), illustrating deterministic temporal scaffolding with minimal stochastic variation.

These observations are informal, and formal perceptual validation remains a target for future work (Section~\ref{sec:future_protocol}). Algorithmic validation (the redundancy and degradation analyses) shows that the pipeline produces output that is structurally distinguishable.

To connect the framework with established information-theoretic analysis of musical structure, we applied a lag-1 redundancy measure (LR) to the canonical instantiation, defined as the mutual information $I(X_t; X_{t-1})$ between consecutive symbolic states such as pitch class. It measures how much the immediately preceding state reduces uncertainty about the present one, and so reflects predictability. This is the first-order case of the redundancy $I(X; Z)$ between the present and the whole past, the quantity that \citet{dubnov2006spectral} calls the information rate and that \citet{abdallah2009information} prefer to call redundancy, reserving \emph{information rate} for their own predictive measure over the future. Restricting the measure to a single lag is a deliberate simplification, and it is exact only under a first-order Markov assumption that the generated sequences are not required to satisfy. Symbol-$A$ sections (deterministic, scale-based pitch and constant IOI) exhibited higher redundancy than symbol-$B$ sections (textural, high-entropy pitch and exponential IOI) in representative runs, consistent with the interpretation that deterministic sections carry more predictable structure.
\subsubsection{Pipeline Component Necessity: Ablation Analysis}
\label{sec:ablation}

The preceding sections demonstrate that the four-layer pipeline preserves the designed distributional separations. A complementary question is whether each layer is \emph{necessary}, that is, whether removing a layer produces detectable structural degradation. Three ablation conditions isolate the contribution of each architectural component (Table~\ref{tab:ablation}).

To show that the L-system string carries structure beyond the ordering of symbols, we compared the grammar-generated sequence (rules $A \rightarrow AB$, $B \rightarrow A$, axiom $A$) with a \emph{shuffled} sequence that preserves the same A:B composition ratio but destroys sequential dependencies. For the symbol-level sequence, we computed the lag-1 redundancy $\mathrm{LR} = I(X_t; X_{t-1})$ (mutual information between consecutive symbols) and the Lempel-Ziv (LZ) parsing complexity (the number of distinct phrases). A higher LR indicates that the previous symbol better predicts the next, and a lower LZ complexity indicates greater compressibility and self-similarity. Table~\ref{tab:lsystem_ir_lz} reports results for expansion depths 4--7, including one-sided permutation $p$-values (1000 shuffles). The ablation results show that the L-system-based macro-form exhibits a \emph{significantly} higher lag-1 redundancy than a random sequence with the same symbol composition (e.g.\ $I(X_t; X_{t-1}) \approx 0.35$--$0.52$ for the L-system vs.\ ${\approx}0.02$--$0.14$ for the shuffled mean; $p_{\mathrm{LR}} \leq .001$ at depths 6--7). A deterministic grammar must be more predictable than permutations of its own output, so these tests are a check that the grammar's structure is measurable at the symbol level rather than a discovery about it. The question they leave open is whether that structure reaches the event stream, which the next comparison addresses. At depths 6--7, the L-system also achieves significantly lower LZ phrase counts than the shuffled baseline ($p_{\mathrm{LZ}} \leq .007$), which corroborates greater compressibility. Thus, the L-system contributes measurable \emph{informational} structure to the macro-form, independent of the distributional metrics applied to event-level output. 

To isolate the effect of recursion-depth modulation on the rendered output, we compared two pipelines on the same symbol sequence ($ABAABABA$, seed~42). These were (i)~depth-weighted, where each symbol's IOI and pitch distributions are modulated by its L-system generation (deeper $\rightarrow$ denser IOI, wider pitch range) and (ii)~symbol-only, where the same sequence is used but generation is ignored so that only the symbol identity drives the mapping. Both outputs were discretised (IOI bins and pitch class), and the LZ phrase count was computed. The comparison is made on the Layer~3 streams, before Layer~4 removes any event. Because depth-weighted sections are denser, total event counts differ (6{,}591 vs.\ 3{,}555), so we report \emph{normalised LZ} (phrases per event). The depth-weighted pipeline yielded a normalised LZ of $0.50$ and the symbol-only pipeline $0.56$. A lower normalised LZ indicates greater compressibility and therefore greater hierarchical self-similarity in the event stream. The comparison is a single pair of runs of unequal length, and normalised LZ retains some length dependence, so the result is suggestive rather than a test. Since the modulation scales all sections of this piece alike, the difference reflects the denser and wider distributions it produces rather than any contrast between sections. 

\begin{figure}[htbp]
\centering
\includegraphics[width=0.95\textwidth]{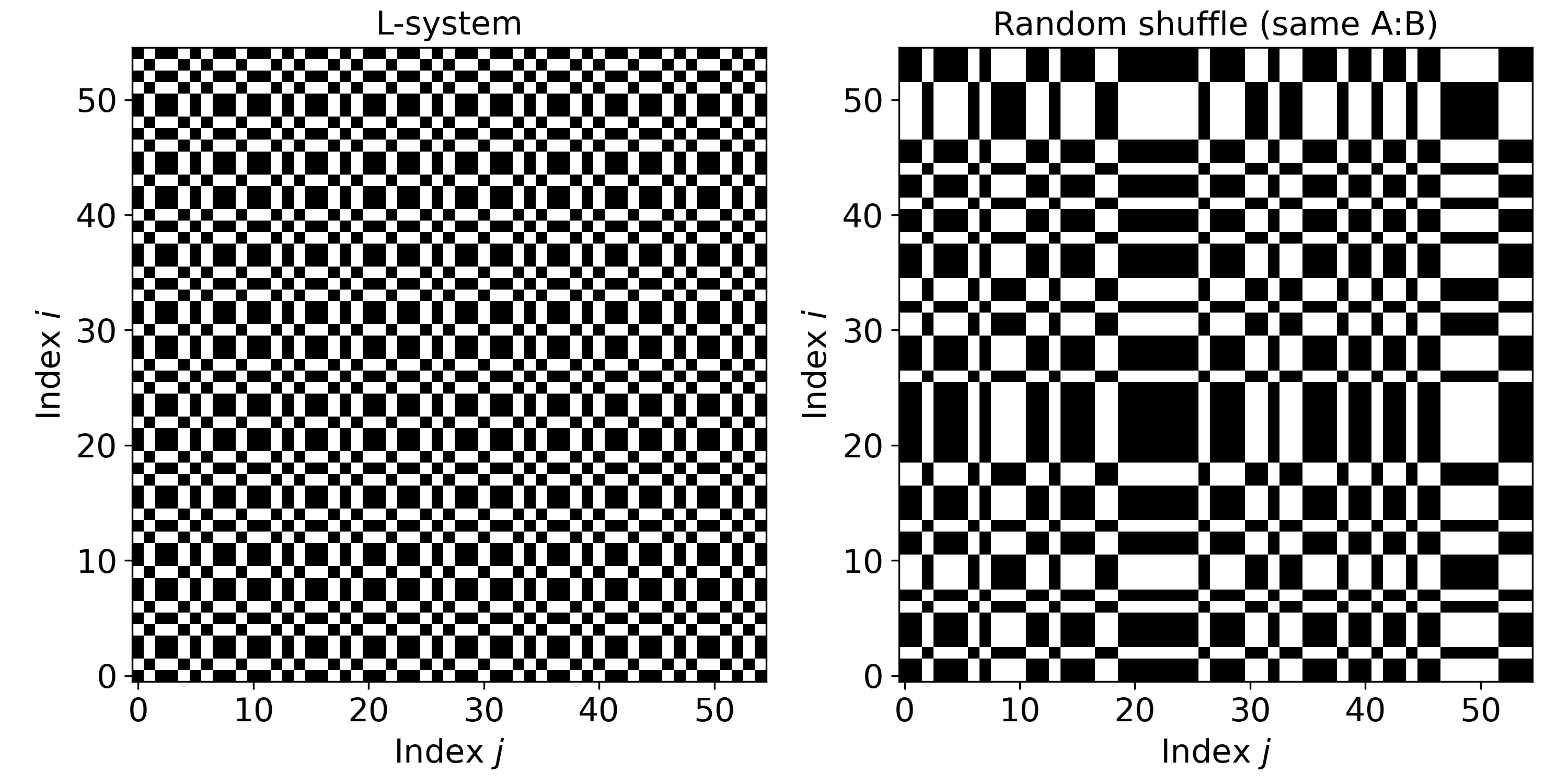}
\caption{Recurrence plots of the L-system sequence (left) and a random shuffle of the same symbols (right), at depth~8.}
\label{fig:recurrence}
\end{figure}

We further quantified the \emph{structural determinism} of the L-system sequence using Recurrence Quantification Analysis (RQA). A recurrence plot records, in a matrix, where the same symbol reappears at two time indices $(i,j)$. Deterministic structure yields extended \emph{diagonal lines} (repeated subsequences), while a random ordering of the same symbols yields fragmented points. The RQA measure \emph{Determinism} (DET) is the proportion of recurrence points that form diagonal lines of length $\geq 2$. We compared the grammar-generated sequence with 500 random shuffles preserving the same A:B composition. Figure~\ref{fig:recurrence} shows the two recurrence plots. The difference between them is one of degree rather than of visual kind, which is why it is reported through DET rather than read off the plot. Table~\ref{tab:lsystem_det} reports DET and a one-sided permutation $p$-value (the proportion of shuffles with DET $\geq$ L-system DET). At depth~8 (string length $|S|=55$), the L-system exhibits significantly higher DET than the random baseline ($p = .032$), the one depth at which the sequence is long enough for the permutation test to resolve the difference. At shallower depths (4, 6), the effect is in the same direction but is not statistically significant with the present sample size. 

\begin{table}[htbp]
\centering
\small
\tbl{Information-theoretic comparison of the L-system and shuffled symbol sequences of the same A:B composition (1000 permutations).}
{%
\resizebox{\linewidth}{!}{%
\begin{tabular}{@{}crrrrrrrr@{}} \toprule
\textbf{Depth} & $|S|$ & \textbf{A:B} & \textbf{LR (L-sys)} & \textbf{LR (Shuffled)} & \textbf{$p_{\mathrm{LR}}$} & \textbf{LZ (L-sys)} & \textbf{LZ (Shuffled)} & \textbf{$p_{\mathrm{LZ}}$} \\ \midrule
4 & 8  & 5:3  & 0.522 & $0.14 \pm 0.17$ & .075 & 5 & $6.0 \pm 0.7$ & .257 \\
5 & 13 & 8:5  & 0.344 & $0.08 \pm 0.11$ & .092 & 6 & $7.7 \pm 0.9$ & .080 \\
6 & 21 & 13:8 & 0.420 & $0.04 \pm 0.06$ & .001 & 7 & $9.9 \pm 0.9$ & .007 \\
7 & 34 & 21:13 & 0.357 & $0.02 \pm 0.03$ & $<$.001 & 8 & $12.6 \pm 1.0$ & $<$.001 \\ \bottomrule
\end{tabular}%
}%
}
\label{tab:lsystem_ir_lz}
\end{table}

\begin{table}[htbp]
\tbl{RQA Determinism (DET) for the L-system and 500 random shuffles of the same A:B composition.}
{\begin{tabular}{@{}crrrrl@{}} \toprule
\textbf{Depth} & $|S|$ & \textbf{DET (L-system)} & \textbf{DET (Random) mean $\pm$ std} & \textbf{$p$-value} & \textbf{Significant} \\ \midrule
4 & 8  & 0.692 & $0.567 \pm 0.104$ & .202 & No \\
6 & 21 & 0.764 & $0.702 \pm 0.038$ & .056 & No \\
8 & 55 & 0.781 & $0.750 \pm 0.016$ & .032 & Yes \\ \bottomrule
\end{tabular}}
\label{tab:lsystem_det}
\end{table}

Ablation~(a) removed the L-system. Layer~1 was replaced by random permutations of the symbol string, preserving the A:B ratio (5:3) but destroying the Fibonacci growth pattern. Although local distributional metrics (MC, RC) remained unchanged ($p > .05$), the informational predictability of the section sequence was degraded. The section-sequence LR fell from $0.52$ for the grammar-generated string to $0.14 \pm 0.17$ across 1000 shuffles, a difference that is directional but does not reach significance at this depth (one-sided permutation $p = .075$; see Table~\ref{tab:lsystem_ir_lz} for the depth series, where the effect is significant from depth 6 onwards). Layer~1 thus functions not as a micro-statistical generator but as a macro-formal regulator that ensures the deterministic evolution of musical states, providing a form of structural coherence that random arrangements fail to preserve.

Ablation~(b) removed the tempo canon. All tempo ratios were set to 1:1 (unison), eliminating the time-scaling that differentiates voices within a section while retaining all other pipeline components. The temporal component of the voice separation score decreased by 78.4\% ($U = 64.0$, $p < .001$, rank-biserial $r = -1.00$; $n = 8$ sections per condition), and same-symbol rhythmic coherence decreased by 18.1\% ($U = 51.0$, $p = .050$, $r = -.59$), an effect that sits at the conventional threshold rather than clearly beyond it. The pitch component of VSS also fell, from 5.45 to 3.34, so the canon's contribution is not confined to the temporal domain. Tempo-canon scaling is thus the primary mechanism for inter-voice temporal differentiation, contributing structure that Layer~2's distributional specification alone cannot provide.

Ablation~(c) removed hardware compensation. Layer~4 pre-compensation was disabled, passing raw Layer~3 timestamps to the output. Under both linear and power-law ($c = 0.5$) latency models, removing compensation introduced systematic velocity-timing coupling, with $r = -1.000$ for the linear model and $r = -.998$ for the power-law model (Table~\ref{tab:ablation}). The onset alignment standard deviation increased from the trivially exact $0.000$~ms to $0.329$~ms (linear) and $0.209$~ms (power-law). As noted in Table~\ref{tab:ablation}, the full-pipeline zero is trivially exact in the simulation. What the ablation demonstrates is the \emph{necessity} of Layer~4. Without compensation, a systematic bias couples velocity to onset timing, distorting the temporal relationships that the preceding layers have constructed. Both model forms give the same coupling, as they must, since without compensation the onset error is a deterministic function of velocity. The correlations verify that the ablation was applied rather than measuring an independent effect.

Ablations~(b) and~(c) confirm that Layers~3 and~4 each contribute measurable, non-redundant structure to the pipeline output. Tempo-canon scaling provides inter-voice temporal differentiation, and hardware compensation removes systematic velocity-timing bias. Ablation~(a) shows that the L-system layer acts at the macro-formal level. Distributional metrics (MC, RC) are unchanged under random symbol order, while the section-sequence lag-1 redundancy falls, directionally at the canonical depth ($p = .075$) and significantly at greater depths, which identifies Layer~1 as the regulator of formal order rather than of local statistics.

Layer~2 is not ablated, and the reason is structural rather than an oversight. Distribution-switching is the mechanism by which every other layer is parameterised, so removing it does not leave a well-formed pipeline to compare against, and no single-regime control was run. The evidence for Layer~2 is therefore positive rather than subtractive. Symbol-conditioned regimes remain statistically separable at the output (Section~\ref{sec:results_ds}), and removing the recursion-depth modulation of the Layer~2 mapping while holding the symbol sequence fixed reduces hierarchical self-similarity in the rendered MIDI (the normalised-LZ comparison above). Whether a fixed-distribution pipeline would perform comparably on some other criterion is a question these experiments do not settle.

\begin{table}[htbp]
\centering
\small
\tbl{Pipeline component ablation: structural effect of removing each layer (unablated symbol-only pipeline, $N = 3{,}383$ events after Layer~4, 74~s). Ablation~(a) uses 100 random permutations for the MC and RC rows and 1000 for the LR row, and (b)--(c) use matched conditions.}
{%
\resizebox{\linewidth}{!}{%
\begin{tabular}{@{}llccl@{}} \toprule
\textbf{Ablation} & \textbf{Metric} & \textbf{Full} & \textbf{Ablated} & \textbf{Test} \\ \midrule
\multicolumn{5}{@{}l}{\textit{(a) No L-system (random symbol order, 5A:3B preserved)}} \\
 & Sequential self-sim.\ (MC) & 0.772 & $0.776 \pm 0.009$ & $z = -0.40$, $p = .692$ \\
 & Same-symbol MC              & 0.765 & $0.774 \pm 0.005$ & $z = -1.83$, $p = .068$ \\
 & Same-symbol RC              & 0.855 & $0.872 \pm 0.017$ & $z = -1.01$, $p = .312$ \\
 & Section-seq.\ LR            & 0.522 & $0.143 \pm 0.167$ & permutation $p = .075$ \\
\midrule
\multicolumn{5}{@{}l}{\textit{(b) No tempo canon (all ratios 1:1; $n = 8$ sections per condition)}}\textsuperscript{a} \\
 & VSS temporal     & $0.283 \pm 0.058$ & $0.070 \pm 0.024$ & $-78.4\%$;\; $U = 64$, $p < .001$, $r = -1.00$ \\
 & RC (same-symbol) & $0.854 \pm 0.032$ & $0.728 \pm 0.119$ & $-18.1\%$;\; $U = 51$, $p = .050$, $r = -.59$ \\
\midrule
\multicolumn{5}{@{}l}{\textit{(c) No hardware compensation\textsuperscript{b}}} \\
 & Onset align.\ SD, linear (ms) & 0.000\textsuperscript{b} & 0.329 & n/a \\
 & Onset align.\ SD, power (ms)  & 0.000\textsuperscript{b} & 0.209 & n/a \\
 & Vel.--timing $r$, linear       & 0.000    & $-1.000$ & n/a \\
 & Vel.--timing $r$, power        & 0.000    & $-.998$ & n/a \\
\bottomrule
\end{tabular}%
}%
}
\tabnote{%
\textsuperscript{a}Percentage changes are computed on the pooled statistic. Means $\pm$ SD and tests use the eight section-level values.\\ \textsuperscript{b}Full-pipeline zeros are trivially exact in simulation, since the same latency model $L(v)$ drives both compensation and alignment measurement.}
\label{tab:ablation}
\end{table}

\subsection{Hardware Compensation}
\label{sec:hardware}

Layer~4 acts as the HAL, bridging algorithmic abstractions and the instrument's mechanical reality. Because latency characteristics vary across instruments, the HAL is an extensible framework for mapping symbolic timing onto the mechanical requirements of the Disklavier key action, not a definitive correction. The velocity command modulates the current pulse driving each key solenoid, and the harder the key is driven the sooner its hammer reaches the string, a relation that \citet{goebl2005piano} fit with a power law in hammer velocity. Parameterising $L(v)$ as a power law follows that functional form, and it is the same asymmetry Yamaha's own prelay function is built to cancel.

The model is offered calibration-ready rather than calibrated. An audit of six public Disklavier-related data sets (Appendix~\ref{app:audit}) found that none pairs MIDI velocity commands with measured acoustic-onset latency, so the 10 to 30~ms envelope and the exponent $c = 0.5$ are a prior, not a measurement, and the results below characterise the compensation logic rather than its fit to any particular instrument. Two properties follow from that stance and are established in the rest of this section. The compensation degrades gracefully when the assumed curve is wrong (Appendix~\ref{app:latency_mismatch}), and the robustness filter provides a fail-safe for instruments that have not been calibrated at all.

The supplementary code includes an onset-alignment protocol for future empirical validation, which compares recordings made with and without pre-compensation through onset detection. The present article reports no results from it.

\subsubsection{Velocity-Dependent Latency Model}

The acoustic onset latency $L(v)$ is modelled as spanning 30~ms (softest) to 10~ms (loudest). Three candidate models share the boundary conditions $L(0) = L_{\max}$ and $L(v_{\max}) = L_{\min}$, and each targets \emph{raw} electromechanical latency with internal compensation disabled:

\begin{align}
    L_{\text{linear}}(v) &= L_{\max} - (L_{\max} - L_{\min}) \cdot \frac{v}{v_{\max}} \label{eq:linear} \\
    L_{\text{power}}(v) &= L_{\max} - (L_{\max} - L_{\min}) \cdot \left(\frac{v}{v_{\max}}\right)^c, \quad c \in (0,1) \label{eq:power} \\
    L_{\text{log}}(v) &= L_{\max} - (L_{\max} - L_{\min}) \cdot \frac{\log(1 + kv/v_{\max})}{\log(1+k)} \label{eq:log}
\end{align}

The maximum inter-model disagreement of $\sim$4.9~ms occurs at mid-velocity ($v \approx 512$), and all models converge at the boundary velocities. On instruments with active \emph{prelay} or AccuPlay compensation, the pre-compensation offset should be replaced by the measured residual latency curve. Pre-compensation shifts each event earlier by its predicted latency, $t_{\text{adjusted}} = t_{\text{intended}} - L(v) / 1000$~[seconds].

\subsubsection{Compensation Performance and Pipeline Integration}

Table~\ref{tab:calibration} summarises the simulated compensation results.

\begin{table}[htbp]
\tbl{Simulated jitter suppression on the software side ($N = 526$ notes, 30-second excerpt).}
{%
\begin{tabular}{@{}lcc@{}} \toprule
\textbf{Condition} & \textbf{Mean abs.\ error (ms)} & \textbf{vs.\ Uncorrected} \\ \midrule
Uncorrected & $17.68 \pm 4.43$ & n/a \\
Linear pre-compensation & $1.27 \pm 0.46$ & $-92.8\%$ \\
Calibrated (power-law, $c=0.5$) & $0.37 \pm 0.23$ & $-97.9\%$ \\ \midrule
Robustness filter (uncalibrated) & $2.24 \pm 2.41$ & $-87.3\%$ \\ \bottomrule
\end{tabular}%
}
\label{tab:calibration}
\end{table}

The simulated error ($0.37 \pm 0.23$~ms) falls below the instrument's ${\sim}$1~ms scanning resolution, effectively eliminating the computational layer as a timing bottleneck. Practical timing remains bounded by the scanning rate. Ablation (Section~\ref{sec:ablation}) confirms that removal of Layer~4 introduces systematic velocity-timing coupling ($|r| > .99$), demonstrating the necessity of a compensation stage independent of absolute precision.

The robustness filter, which compresses flagged velocities towards the local mean in a 50~ms window, reduces error variability in the uncalibrated case (SD of the absolute error from 4.24~ms to 2.87~ms in the filter's distorted-against-corrected comparison, a reduction of 32.4\%, $p < .001$), but \emph{degrades} calibrated performance by invalidating velocity-matched corrections (mean error increases from 0.37 to 2.24~ms). The robustness filter is a fail-safe for uncalibrated deployment, and under the simulated uncalibrated condition it retains a jitter reduction of 87.3\% ($p < .001$) relative to the raw MIDI output.

Sensitivity analysis (Appendix~\ref{app:powerlaw_sensitivity}) shows that residual jitter after compensation is minimal at $c = 0.5$ and increases smoothly when the exponent deviates (e.g.\ 1.04~ms at $c = 0.3$, 0.74~ms at $c = 0.7$), so the calibrated choice is not tied to a single parameter. A latency model mismatch experiment (Appendix~\ref{app:latency_mismatch}) simulates the case where the true piano deviates from the assumed power-law, either through an exponent $c_{\mathrm{true}} \in [0.3, 0.7]$ or through additive per-note noise up to $\pm 2$~ms. For example, at $c_{\mathrm{true}} = 0.3$ the uncorrected jitter is 2.47~ms against 1.04~ms with the HAL, and at $\pm 2$~ms noise it is 3.69~ms against 1.15~ms. Applying the correction is always preferable to applying none, even when the model is imperfect.

\begin{figure}[htbp]
\centering
\includegraphics[width=0.78\textwidth]{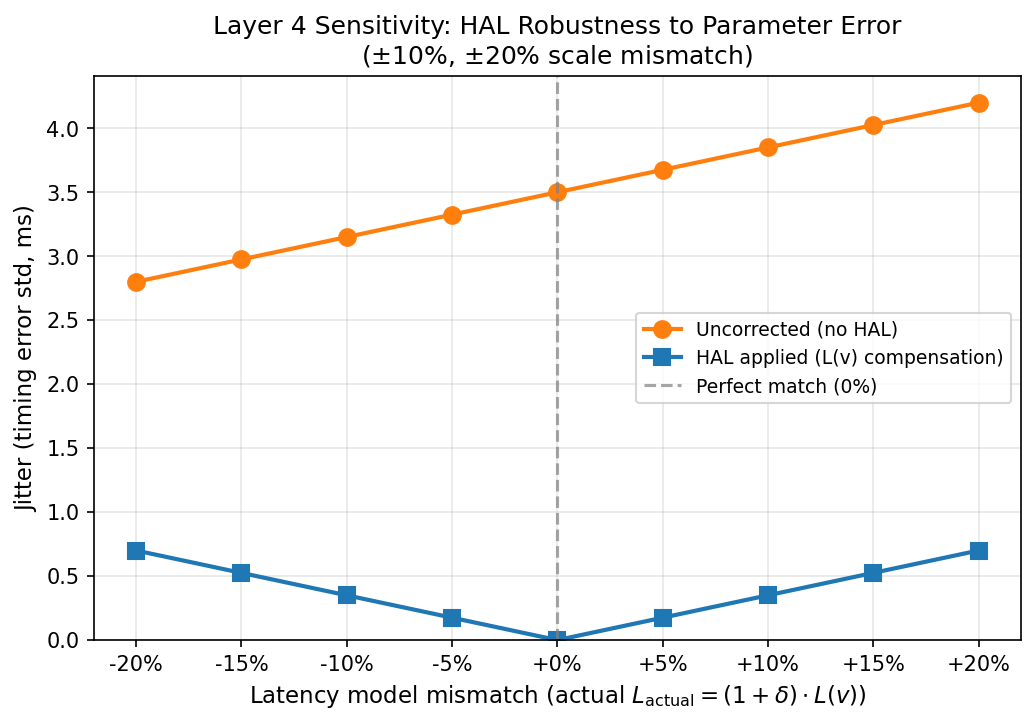}
\caption{Sensitivity of the Layer~4 HAL to latency model mismatch, for a scale error $\delta$ from $-20\%$ to $+20\%$ ($N = 526$ notes).}
\label{fig:latency_sensitivity}
\end{figure}

To characterise robustness to real-world calibration error, we performed a sensitivity analysis in which the assumed latency model was deliberately mismatched, with the actual latency set to $L_{\mathrm{actual}}(v) = (1 + \delta)\,L(v)$ with $\delta \in [-20\%, +20\%]$ (scale error). Figure~\ref{fig:latency_sensitivity} shows that even under a 10--20\% model mismatch, jitter with HAL applied remains well below the uncorrected jitter. In the extreme case of $\pm 20\%$, HAL preserves a jitter reduction of more than 70\% compared with no correction (e.g.\ uncorrected ${\approx}$2.8--4.2~ms vs.\ with HAL ${\approx}$0.7~ms). The compensation therefore remains effective when the true latency is a rescaling of the nominal model by up to $\pm 20\%$. Whether it holds for a curve of a different shape is not tested here. A complementary ``virtual real piano'' simulation (Appendix~\ref{app:latency_mismatch}) models a piano whose latency follows the nominal power-law with $\pm 10\%$ per-note noise and parameter drift. Over 200 trials ($N = 526$ notes), the onset jitter was (A) $3.63 \pm 0.10$~ms (no correction), (B) $0.93 \pm 0.03$~ms (HAL, $c = 0.5$), and (C) 0~ms (ideal correction). The paired comparison of (A) and (B) showed that HAL produces significantly lower jitter ($p < .001$), which indicates that the power-law assumption alone provides substantial benefit without perfect calibration (Table~\ref{tab:virtual_real_piano}).

The interaction between Layer~4 compensation and the preceding layers is quantified in the degradation analysis (Table~\ref{tab:degradation}). Layer~4 leaves the velocity distribution itself unchanged, and the effect of compensation appears in the IOI row for symbol $B$, where the KS distance rises from 0.032 to 0.322, because pre-compensation ties each onset to its velocity and so creates a timing-velocity coupling absent from the original event stream. This coupling leaves the pitch and velocity fidelity metrics untouched, but it is relevant for compositions in which velocity-timing independence is a design goal.

\subsection{Computational Coherence Metrics Across Density}
\label{sec:coherence}

This section characterises the density range over which coherence metrics retain sensitivity and the transition above which alternative strategies are required.

\subsubsection{Metric Behaviour Across Density}
\label{sec:coherence:breakpoint}

The question this section asks is at what density a single-domain melodic metric stops being able to tell structured music apart from noise. We operationalise it directly, as the separability of structured streams from a random null under a fixed temporal integration window, and we measure it rather than assume it.

The measurement models temporal resolution with an integration window $w$, within which two onsets are treated as unresolved and merge into a single unit. The window is a modelling assumption, not a measured perceptual limit. We fix $w = 40$~ms, a resolution rate of 25~Hz that lies just above the fusion rate of roughly 20~events/s discussed in Section~\ref{sec:bg_asa}, and the sensitivity of the result to this choice is checked below. At each of fourteen aggregate densities from 10 to 200~notes/s the system's generator produces a structured melodic stream (real symbol configurations, C-major pitch set) and a random null stream. Both are collapsed by the integration window, and we measure the fraction of the intended melodic transitions that survive, which we call \emph{retention}. Discriminability is the retention of the structured stream minus that of the null. The procedure runs 40 trials per density and is reproducible with a single script in the accompanying code repository.

Retention of the structured stream falls smoothly with density, from $0.925 \pm 0.020$ at 10~notes/s to $0.248 \pm 0.008$ at 200~notes/s, as more events merge inside each window. This is a genuine density dependence, and it stands in contrast to a fixed-count pitch-interval entropy such as SVC, which cannot vary with density at all. The behaviour of interest, however, is not in the structured curve alone but in its separation from the null, which is plotted in Figure~\ref{fig:melodic_threshold}.

Discriminability is greatest at 25~notes/s, where the structured stream retains $0.857 \pm 0.028$ of its transitions against the null's $0.635 \pm 0.027$, a gap of $0.222$ (Welch $t(78) = 36.0$, $p < .001$, $d = 8.06$). The window's 25~Hz resolution rate and the 25~notes/s peak coincide numerically, and the window sweep reported below shows that the peak does not move with $w$, so the coincidence is not what places it. The peak is stable, in that a bootstrap over trials returns 25~notes/s in every resample (95\% CI $[25, 25]$, a resolution bounded by the 14-level grid, whose neighbouring levels are 20 and 28~notes/s). Above it the gap collapses, falling below half its peak value by 40~notes/s and reaching a minimum of $0.034$ at 80~notes/s. The peak sits just above the rate of roughly 20~events/s at which discrete events are reported to fuse into texture, and the collapse just above it is the loss of single-domain discriminative power the paper set out to locate \citep{roads2001microsound, vannoorden1975temporal, bregman1990auditory}.

The gap does not vanish above the band, and the reason is instructive. It recovers at the highest densities, to $0.125$ at 200~notes/s ($d = 14.0$). By that density the structured stream has lost its melodic ordering to integration just as the null has, but it has not lost its \emph{pitch-class content}. Its onsets still fall on the seven classes of its scale while the null's spread across all twelve. The two streams differ in pitch set as well as in ordering, so the recovered gap is consistent with pitch-class content surviving integration when ordering does not, although separating the two contributions requires a measurement these experiments do not provide. Melodic discriminability is therefore not lost above a threshold once and for all. It is lost within a band, roughly 40 to 100~notes/s, above which what distinguishes structured from random music is no longer its melodic contour but its distributional content. This is the same conclusion the cross-domain constraint analysis reaches from the other direction (Section~\ref{sec:crossdomain}), and it is why distributional content, rather than melodic order, becomes the effective lever at extreme density.

\begin{figure}[htbp]
\centering
\begin{tikzpicture}
\begin{axis}[
    width=0.92\textwidth, height=0.47\textwidth,
    xlabel={Aggregate density (notes/s)}, ylabel={Transition retention},
    xmin=5, xmax=210, ymin=0, ymax=1.0,
    grid=major, grid style={gray!20},
    font=\small, legend pos=north east
]
\addplot[blue, thick, mark=*, mark size=1.4] coordinates {
    (10,0.925) (15,0.901) (20,0.873) (25,0.857) (28,0.781) (30,0.737) (40,0.592) (50,0.503) (60,0.430) (80,0.333) (100,0.282) (120,0.275) (150,0.268) (200,0.248)
};
\addplot[orange, thick, mark=triangle*, mark size=1.4] coordinates {
    (10,0.821) (15,0.752) (20,0.687) (25,0.635) (28,0.607) (30,0.580) (40,0.505) (50,0.432) (60,0.384) (80,0.299) (100,0.241) (120,0.208) (150,0.163) (200,0.122)
};
\addplot[red, thick, mark=square*, mark size=1.2, dashed] coordinates {
    (10,0.104) (15,0.149) (20,0.186) (25,0.222) (28,0.174) (30,0.157) (40,0.087) (50,0.071) (60,0.046) (80,0.034) (100,0.041) (120,0.067) (150,0.105) (200,0.125)
};
\draw[dotted, thick, gray] (axis cs:25,0) -- (axis cs:25,0.95);
\node[font=\footnotesize, anchor=west] at (axis cs:48,0.72) {peak separability: 25 notes/s};
\draw[gray, -{Stealth[length=2mm]}] (axis cs:47,0.72) -- (axis cs:26,0.26);
\legend{Structured, Random null, Discriminability}
\end{axis}
\end{tikzpicture}
\caption{Melodic-transition retention under a 40~ms integration window against density (40 trials per level). The series are the structured stream, the random null, and their difference (dashed).}
\label{fig:melodic_threshold}
\end{figure}

The location of peak separability is a property of the measurement, and we tested whether it is set by the integration window as the mechanism would require. Varying $w$ from 20 to 50~ms leaves the qualitative picture unchanged (the peak stays in the 25~notes/s region), but the piecewise break point of the retention curve does not track $1/w$. It sits near 60~notes/s across windows, which tells us the knee in the structured curve is where retention approaches its floor on this density grid, not a temporal-resolution point. We therefore do not report a single sharp threshold. What is stable is the density of maximal separability and the band over which separability is lost, not a knife-edge above which coherence disappears.

\subsubsection{Cross-Domain Constraint Engineering}
\label{sec:crossdomain}

Table~\ref{tab:constraints_results} reports the constraint conditions described in Section~\ref{sec:eval_framework}, measured in raw VSS. A registral constraint alone raises PCC but leaves VSS essentially unchanged, since the voices still share their velocity and timing profiles. Adding disjoint velocity bands and distinct temporal profiles raises VSS from 3.65 to 71.08, an increase the component analysis of Section~\ref{sec:wvss_validation} attributes almost entirely to the velocity bands. At high density, measurable separation is therefore gained across domains rather than within the pitch domain alone.

\begin{table}[htbp]
\tbl{Cross-domain constraint effects on voice separation ($N = 4$ voices, 30 realisations per condition).}
{\begin{tabular}{@{}lccl@{}} \toprule
\textbf{Constraint} & \textbf{PCC Change} & \textbf{VSS Change} & \textbf{Significance} \\ \midrule
None (baseline) & n/a & 3.65 & n/a \\
Pitch ($> 5$ st.) & $+554.8\%$ & Minimal & $p < .001$ \\
Pitch + Temporal + Velocity & $+549.7\%$ & $+1849\%$ & $p < .001$ \\
\quad (stratified bands) & & (71.08) & $H = 12.32$** \\ \bottomrule
\end{tabular}}
\tabnote{**$p = .002$. The $H$ statistic is the Kruskal-Wallis test on VSS across the three conditions, computed over the six voice pairs of each condition. The PCC change is tested separately ($H = 61.60$, $p < .001$).}
\label{tab:constraints_results}
\end{table}

\subsubsection{Weighted Voice Separation Score and Component Analysis}
\label{sec:wvss_validation}

Component analysis of raw wVSS reveals an extreme dominance of velocity contrast within the tested condition, with $w_{\text{vel}} = 99.96\%$ against $w_{\text{pitch}} = 0.02\%$ and $w_{\text{temporal}} = 0.01\%$ ($N = 4$ voices, high-density condition). This raw dominance has two sources, a \emph{scale effect} (velocity spans 0--1023 XPMIDI units vs.\ 0--127 MIDI notes for pitch) and a \emph{stratification effect} (voices were assigned distinct velocity bands by design). Because the tested condition intentionally maximises velocity-band separation, the raw wVSS figure of 99.96\% represents an upper bound on the effectiveness of this strategy rather than a general property of the metric.

To disentangle the scale from stratification, we use the range-normalised variant nwVSS (Equation~\ref{eq:nwvss}) as the primary cross-domain comparison metric in the remainder of this section. Range normalisation barely moves it. Velocity still accounts for 98.3\% of nwVSS in the same condition, with pitch at 0.2\% and the temporal domain at 1.5\%. The dominance therefore does not dissolve once the scale advantage is removed, which is the useful finding. Dynamic stratification is the most efficient single lever for maintaining measurable inter-voice differentiation at high density, and that is a property of the texture rather than an artefact of the units in which velocity happens to be measured. The cross-domain comparisons that follow are reported in nwVSS unless otherwise noted.

To ensure the objectivity of the wVSS and mitigate potential circular reasoning (weights being derived from the same data to which they are applied), we conducted split-half cross-validation. The 500-event data set was divided into two halves, indexing events by even and odd position. The weights derived from the even-indexed half ($w_{\text{vel}} = 99.96\%$) agree with those from the odd-indexed half ($w_{\text{vel}} = 99.97\%$), with a Pearson correlation between the two weight vectors of $r > .99$. The maximum weight deviation between splits is 0.007 percentage points for wVSS and 0.055 for nwVSS, rising to 0.52 for nwVSS under 100 random 50:50 splits. This indicates that the observed dominance of the dynamic domain at high densities is a structural property of the generated texture rather than a sample-specific bias. In ultra-high-density conditions, dynamics consistently acts as the primary discriminative mechanism for voice separation.

The nwVSS weights also transfer across density (Table~\ref{tab:nwvss_density}). At 20~notes/s, the temporal component accounts for 7.63\% of nwVSS (vs.\ 0.00\% at 120~notes/s), and pitch increases from 0.27\% to 0.45\%. The disappearance of temporal weight at high density is not a metric failure but a computational signature of temporal fusion. At 120~notes/s, the fine structure of the onsets ceases to carry discriminative information for voice separation, consistent with the microsound literature's description of aggregate-texture perception at high event rates \citep{roads2001microsound, mcdermott2013summary}. This weight transfer reinforces the operational-envelope finding of Section~\ref{sec:coherence}. As density rises through the loss band, the relative contribution of pitch and temporal domains to measurable separation decreases, shifting the discriminative burden to dynamics.

A final variant couples velocity to register explicitly, through $v = 200 + 12.5 \times (p - 40)$ over MIDI notes 40 to 105. The near-unit correlation between pitch and velocity that follows ($r = .999999$) is not a finding but a check that the constraint was applied as specified, since velocity is a deterministic function of pitch by construction. What the variant demonstrates is that coupling the two domains raises wVSS far above any single-domain constraint ($U = 0.0$, $p = .002$), which is the compositional point. A composer who wants voices to remain separable at extreme density gains more from tying dynamics to register than from separating either domain alone.

To complement the VSS with a measure independent of dynamics, we define the Pitch Class Set (PCS) distance between a pair of voices. Within 1-second windows, each voice is represented by a 12-dimensional pitch-class histogram, and the distance is defined as $1 - \text{cosine similarity}$ between these vectors. The PCS distance quantifies separation in \emph{harmonic territory} (which pitch classes each voice occupies) without conflating it with the velocity-band separation. We report it here as the measure a reader should ask for, rather than as a result. The constraint conditions of this section were designed around velocity stratification, and we have not run PCS distance over them, so the claim that the voices are separated in harmonic territory as well as in dynamics is one the present experiments do not establish. It is the obvious next measurement, and the metric is implemented in the supplementary code.

Constraint ordering also affects generation cost. Three strategies were compared over 10 replications each, using an efficiency index defined for this comparison as the wVSS achieved divided by one plus the percentage of candidate events rejected. Chained reactive application (pitch$\rightarrow$velocity$\rightarrow$temporal) was $52.6\%$ more efficient than simple reactive application (Welch $t$-test, $p = .026$) and $32.7\times$ more efficient than global simultaneous rejection ($p < .001$), which discards almost every candidate before one satisfies all constraints at once.

\subsection{Convergence Point Calculus as Distribution-Switching Trigger}
\label{sec:convergence}

Convergence-point triggering provides a principled interface between the deterministic temporal structure of tempo canons (Layer~3) and the parametric layer (Layer~2) that governs distribution-switching. Rather than relying solely on the L-system macro-form to determine when distributional regimes change, CP events enable distribution switches to be triggered by the internal temporal logic of the canon itself, so that when voices converge within tolerance $\epsilon$ the system can switch to a new distributional regime. This feedback loop enables a structural relief mechanism. As deterministic canon complexity reaches its peak at convergence, the system triggers a switch to stochastic regimes, resolving temporal density into texture.

\subsubsection{Discrete Event Triggering}

The run has three voices, namely two canon voices in a 3:4 rational relation (IOI $= 1.000$~s and $0.750$~s) and one Poisson voice, over 60~seconds. The distribution switch is activated at the convergence at $t = 30$~s, so that the first half of the run is pre-CP and the second half post-CP. The statistical tests compare the Poisson voice, the one the switch acts on, across $n = 30$ one-second analysis windows per condition. Excerpt~4 is a 30-second rendering of the same design, with the switch at $t = 15$~s.

\begin{table}[htbp]
\tbl{Convergence-point triggered parameter switch in the stochastic voice, pre-CP against post-CP ($n = 30$ one-second windows per condition).}
{\begin{tabular}{@{}lccl@{}} \toprule
\textbf{Parameter} & \textbf{Pre-CP} & \textbf{Post-CP} & \textbf{Test} \\ \midrule
Poisson-voice density (notes/s) & $5.33 \pm 1.89$ & $38.27 \pm 6.30$ & $U=0$, $r = 1.000$ \\
Pitch-Class Concentration & $0.523 \pm 0.141$ & $0.069 \pm 0.035$ & $U=900$, $r = -1.000$ \\ \bottomrule
\end{tabular}}
\tabnote{Both $p < .001$. With $n = 30$ per condition, max $U = 900$.}
\label{tab:cp_discrete}
\end{table}

The switch is visible in both parameters (Table~\ref{tab:cp_discrete}). Both statistics verify that the switch fired where it was armed rather than reporting an independent effect. The post-CP density of 38.27~notes/s sits at the lower edge of the 40--100~notes/s band over which melodic metrics lose their discriminative power (Section~\ref{sec:coherence}), moving the post-switch texture out of the region where melodic order is the discriminating feature. A convergence point can thus be positioned to move the music into that band, letting the canon's temporal logic control which domain carries the structure.

\subsubsection{Continuous Parameter Modulation}

An $e{:}\pi$ tempo canon with its convergence target at $t_{CP} = 15$~s ($\epsilon = 50$~ms) drives an inhomogeneous Poisson voice at rate $\lambda(t) = 5 + 40 \cdot |t - t_{CP}| / t_{CP}$ over 30~s ($n = 761$ events under seed 42).

The achieved per-window density tracks the intended rate with Pearson $r = .928$ ($p < .001$), root-mean-square error $= 4.70$~notes/s, and the tracking holds on both sides of the convergence (pre-CP $r = .896$, post-CP $r = .961$). Near the convergence (12--18~s) the density falls to 10.2~notes/s against 40.5~notes/s at the extremes, a $4.0\times$ reduction.

The continuous modulation demonstrates that the CP calculus can serve as a smooth control interface, not only for discrete switching but also for gradual parameter evolution. The parameter $\epsilon$ is not an arbitrary detection constant, but a \emph{convergence resolution} parameter that mediates between sharp convergence events (small~$\epsilon$, abrupt distribution switches) and diffuse transitions (large~$\epsilon$, gradual modulation). The default value $\epsilon = 50$~ms is hardware-aware. It matches the key reset time (Table~\ref{tab:hardware}), below which a convergence falling on the same key would be a \emph{ghost convergence}, detected in software but not physically realisable because the action has not reset. Other values remain available as a compositional choice.

The sensitivity analysis (Appendix~\ref{app:epsilon_sensitivity}) confirms that the behaviour of the system is regular over $\epsilon \in \{10, 20, 50, 100\}$~ms, invariant for the rational canon and monotonic for the irrational one. For the rational 3:4 canon, the convergence count is invariant (11 events in 30~s in all tested $\epsilon$), reflecting the exact periodicity of rational ratios. For the irrational canon $e:\pi$, the count increases monotonically with $\epsilon$ (5 at 10~ms to 51 at 100~ms; Table~\ref{tab:epsilon_sensitivity}), providing the composer with continuous control over the switching density. This monotonic relationship is compositionally desirable, since an increase in $\epsilon$ smoothly increases the frequency of distribution-switching events, enabling fine-grained control over textural volatility without introducing discontinuities or bifurcations in system behaviour.

\section{Discussion}
\label{sec:discussion}

\subsection{Theoretical Contributions}

This work demonstrates that historically separate compositional methodologies can cohabit within the four-layer architecture through hierarchical distribution-switching. The integration is architectural rather than theoretical. The framework does not claim that these traditions reduce to a common formalism, but rather that distribution-switching provides a practical interface through which they can cohabit and interact within the same event-generation process. Qualitative musical differentiation arises from switching distribution \emph{types} (constant vs.\ exponential IOI, low vs.\ high-entropy pitch sets, constant vs.\ variable velocity) rather than from modulating parameters within a fixed family. These designed differences survive the pipeline with the effect sizes reported in Table~\ref{tab:coherence}. The degradation analysis (Table~\ref{tab:degradation}) locates the greatest per-layer distortion in IOI, at Layer~4, where latency compensation acts only where velocity varies and the key-reset mask acts wherever a key is re-struck within 50~ms, while pitch is transmitted with near-zero loss.

Ablation~(a) shows that the L-system acts on the ordering and predictability of section transitions rather than on the note distributions within a section, which are unchanged under random symbol order. At the depth used for the canonical piece the effect is directional but not significant, since a depth-4 expansion yields only eight symbols and a permutation test on so short a sequence has little power (LR $0.52$ against $0.14 \pm 0.17$ for the shuffled null, $p = .075$). It becomes significant once the sequence is long enough to carry the structure, at depth 6 ($p = .001$) and depth 7 ($p < .001$). The grammar therefore encodes a macro-formal state-space whose information-theoretic signature is absent from random permutations of the same symbol counts, and the canonical piece is simply too short a sample of it to demonstrate that on its own.

The convergence point calculus bridges the deterministic and stochastic domains, in that $\epsilon$ controls convergence resolution from sharp switching (small $\epsilon$) to gradual modulation (large $\epsilon$), with monotonic behaviour across $\epsilon \in [1, 100]$~ms (Appendix~\ref{app:epsilon_sensitivity}).
The CP feedback path (Figure~\ref{fig:architecture}) enables the temporal logic of the canon to modulate the parametric layer, creating a bi-directional relationship between Layers~2 and~3 that neither L-system sequencing nor stochastic generation alone can produce. Collectively, the four contributions exercise distinct inter-layer interfaces and show that $\Theta_s$ survives transmission to the MIDI output, with the IOI distribution at Layer~4 as the one measured exception and the 8.5\% of events suppressed by the key-reset mask as its stated cost.

\subsection{Relationship to Existing Work}

The framework extends Nancarrow's practice \citep{gann1995music} by adding stochastic microstructure within deterministic scaffolds. It extends Xenakis' methods \citep{xenakis1992formalized} by placing his stochastic generators under a grammar-driven macro-form rather than the single-level probabilistic matrix of \emph{Achorripsis}, and it extends L-system composition \citep{manousakis2006musical} by connecting grammar symbols to physically constrained rendering with quantified coherence metrics.

\subsubsection{Systematic Comparison with JCMS Frameworks}

Table~\ref{tab:systematic_comparison} contrasts representative JCMS approaches with the present framework. \citet{lattner2018imposing} impose higher-level structure on polyphonic output, including self-similarity, metre, and tonality, by combining a convolutional restricted Boltzmann machine with gradient-descent constraint optimisation, and their constraints are musical rather than physical. \citet{kaliakatsos2016learning} learn and generate harmonies across diverse idioms with an adaptive modular system that admits user-defined chord constraints. Both address creativity and formalisation, and neither integrates actuator constraints or a physical correction layer, since neither targets an automated instrument. Amanous is distinguished by two contributions that the table highlights. The first is hardware constraint enforcement, in which explicit inequalities (velocity range, per-key rate, polyphony, latency bounds, scanning resolution) are part of the pipeline rather than post hoc checks, with the per-key rate enforced on every emitted trigger by the key-reset mask. The second is the physical correction layer, in which a dedicated Layer~4 formalises velocity-dependent latency and applies pre-compensation so that the numeric layer's output is corrected before actuation. Neither Lattner et al.\ nor Kaliakatsos-Papakostas et al.\ target automated piano hardware or beyond-human density rendering, and Amanous provides a unified pipeline from symbol to actuation-ready MIDI with both constraint enforcement and physical correction.

\begin{table}[htbp]
\tbl{Systematic comparison with representative JCMS frameworks. Hardware-related rows reflect a difference in scope, not a deficiency in the other systems.}
{\begin{tabular}{@{}>{\raggedright\arraybackslash}p{2.9cm}>{\raggedright\arraybackslash}p{2.8cm}>{\raggedright\arraybackslash}p{3.2cm}>{\raggedright\arraybackslash}p{3.75cm}@{}} \toprule
\textbf{Criterion} & \textbf{Lattner et al.} & \textbf{Kaliakatsos-P. et al.} & \textbf{Amanous (this work)} \\ \midrule
L-system / grammar & n/a & Optional & Yes (Layer~1) \\
Tempo canon & n/a & n/a & Yes (Layer~3) \\
Stochastic distributions & n/a & Optional & Yes (Layer~2) \\
Distribution-switching & n/a & n/a & Yes (symbol $\mapsto$ regime) \\
Hardware constraints & No (not targeted) & No (not targeted) & Yes (Table~\ref{tab:constraint_ineq}) \\
Hardware abstraction layer & No (not targeted) & No (not targeted) & Yes (Layer~4) \\
Actuation-ready output & No & No & Yes (Direct MIDI/XPMIDI) \\
Beyond-human density & No & No & Yes (Tested to 200 notes/s) \\
Disklavier hardware & No & No & Yes \\ \bottomrule
\end{tabular}}
\label{tab:systematic_comparison}
\end{table}

\citet{collins2018there} identifies large-scale algorithmic generation as a new direction for the field, and Amanous realises one concrete instance by integrating all three traditions within a hardware-aware pipeline with statistical output validation. Unlike Essl's \emph{Lexikon-Sonate} \citep{essl2017algorithmic}, which uses real-time interaction as an organising principle, our framework ensures exact reproducibility while retaining stochastic variation. Ablinger's \emph{Quadraturen~III} \citep{ablinger2004quadraturen} and Ritsch's \emph{Autoklavierspieler} \citep{ritsch2011robotic} compensated for hardware behaviour per key and per piece, tied to a purpose-built instrument. Amanous instead offers a general latency formalisation that any Disklavier can be calibrated into.

\subsection{Implications for Creative Music Systems}
\label{sec:creativity}

Amanous functions as an augmented compositional instrument rather than an autonomous generator. The creative agency resides in the iterative loop between the composer and the hierarchical pipeline, in which the composer modulates $\epsilon$ and $\Theta_s$ based on metric feedback after each generation (PCC, VSS), working within the ``textural zone'' by tuning cross-domain constraints. This interaction transforms the generation process into a discovery-driven exploration of superhuman note rates while preserving full control over every structural parameter. Within \citet{boden2004creative}'s taxonomy, it supports \emph{exploratory creativity} (systematic traversal of the configuration space $\{\Theta_s\}_{s \in \Sigma}$) and \emph{combinational creativity} (juxtaposition of tempo-canon, stochastic, and L-system concepts within a single pipeline). For example, replacing Symbol~$B$'s exponential IOI with a Gaussian ($\mu = 0.025$~s, $\sigma = 0.008$~s) produced a denser but more uniform texture. The exponential was selected for the canonical instantiation because its temporal clustering created greater contrast with the deterministic $A$ sections, a decision made through iterative listening to rendered output. The framework's main compositional affordance is modularity, since changing a single symbol-to-distribution mapping or grammar rule propagates through the pipeline to restructure the entire work.

Because the system positions creative agency in the composer's selection of $\Theta_s$ rather than in autonomous generation, evaluation frameworks targeting creative agents (e.g.\ SPECS; \citealp{jordanous2012standardised}) would need to assess the human-plus-system composite, an experimental design outside the present scope. The pipeline fidelity and ablation analyses reported above serve a related but distinct function. They validate the system's reliability as an engineering tool, confirming that design intent propagates from grammar symbol to the MIDI output with quantifiable, per-layer degradation.

\subsection{Practical Implications}

The measured separability profile acts as a compositional aid. It marks where the composer's lever must shift from melodic counterpoint to statistical texture. Structured and random streams are most separable near 25~notes/s, and above the 40--100~notes/s band melodic ordering no longer distinguishes them, and what remains detectable is distributional content. Knowing where that shift occurs lets the composer choose the domain that will actually register at a given density, rather than working blind at the limits of temporal resolution. Dynamic stratification remains the most efficient lever for inter-voice differentiation at extreme densities (98.3\% of nwVSS; Section~\ref{sec:wvss_validation}).

The degradation analysis (Section~\ref{sec:layer_analysis}) turns this into actionable guidance. Pitch and velocity switches reach the instrument essentially untouched, whereas an IOI switch is reshaped by Layer~4, by latency compensation in proportion to how much the section's velocities vary and by the key-reset mask in proportion to how often a key is re-struck. A composer who wants a predictable, faithfully transmitted contrast should therefore switch pitch or velocity, and a composer who wants the mechanism's own timing to colour the result should switch IOI.

The density-conditioned weights of the nwVSS (Table~\ref{tab:nwvss_density}) further confirm this density-dependent weight transfer, with the vanishing temporal weight at 120~notes/s reflecting the shift to aggregate-texture representation \citep{roads2001microsound}.

\subsection{Limitations}

\begin{enumerate}
    \item \textbf{Only computational metrics.} All findings are algorithmic. Perceptual validation is left to the protocol proposed in Section~\ref{sec:future_protocol}.

    \item \textbf{The latency model is uncalibrated, and the Layer~4 results cannot calibrate it.} The alignment error of $0.37 \pm 0.23$~ms is an algorithmic figure, and the practical precision is bounded by the instrument's ${\sim}$1~ms scanning resolution. More importantly, every Layer~4 experiment reported here draws the instrument's latency and the compensator's correction from the same parameterised family, so these experiments test whether the compensation logic inverts its own model, which it does, and they cannot test whether that model describes a real Disklavier. They are insensitive by construction to the shape of $L(v)$, and even to its direction. The direction we assume rests on the mechanics of the piano action and on the design of Yamaha's prelay function \citep{goebl2005piano, goebl2003measurement}, not on any measurement of our own, and the only published Disklavier reproduction study measures a residual against a human performance rather than the raw curve \citep{goebl2003measurement}. Settling this requires the velocity sweep proposed in Appendix~\ref{app:audit}, namely commanded velocities across the full range with acoustic onsets captured at sub-millisecond resolution on the target instrument. Until then the model's value is its readiness for calibration rather than its fidelity to any particular piano.

    \item \textbf{Weighting scheme.} Both wVSS and its range-normalised variant nwVSS are reported. Velocity dominance (99.96\% in wVSS) was derived from a four-voice high-density condition and validated by split-half cross-validation (Section~\ref{sec:wvss_validation}). nwVSS provides a check against scale-driven dominance. The nwVSS weights vary more across splits (0.055 percentage points, and 0.52 under random 50:50 splits) than the wVSS weights (0.007), so the normalised weights are the more sample-sensitive of the two. Generalisation to other densities, voice counts, or pitch distributions remains open.

    \item \textbf{Scope of the evidence.} The canonical results come from a single seeded run of one eight-section piece, the group sizes are five and three, and the contrasts tested are ones the configuration specifies, so the reported effect sizes measure how cleanly a designed separation survives the pipeline rather than the discovery of a difference. Layer~2 is not ablated, for the reason given in Section~\ref{sec:ablation}, and no comparison against an alternative generative system was run.

    \item \textbf{The key-reset mask removes events.} Enforcing the 50~ms reset time costs the canonical piece 560 of its 6{,}591 events, and the suppressed events are dropped rather than reassigned to neighbouring keys. A voice-allocation stage that redistributed them would preserve density at the price of altering the pitch distribution, and it is not implemented. Simultaneous actuation limits arising from solenoid current draw are likewise not modelled, so the polyphony bound is the key count alone.

    \item \textbf{Platform specificity.} The framework targets the Yamaha Disklavier. Adaptation to other automated pianos requires characterising platform-specific latency curves.
\end{enumerate}

\subsection{Future Work: Psychoacoustic Validation and Composer Study}
\label{sec:future_protocol}

The density of maximal computational separability, near 25~notes/s, serves as a predictive hypothesis for auditory scene analysis at extreme densities. We propose it as a formal hypothesis for future psychoacoustic validation, predicting that the perceptual inflection point from ``melody'' to ``texture'' will correlate significantly with this computationally derived boundary. Listeners presented with stochastic two-voice textures across the density levels of Section~\ref{sec:coherence} will report a qualitative shift from perceiving individual melodies to perceiving aggregate texture near this density.

We propose a within-subjects design with $N = 30$ participants and the 14 density conditions of Section~\ref{sec:coherence}. In each trial, participants hear a 5-second excerpt and make a forced-choice response (``I hear distinct melodies'' vs.\ ``I hear a texture''). The resulting psychometric function, that is, the proportion of ``texture'' responses as a function of density, should exhibit a sigmoidal transition whose inflection point can be compared against the measured separability peak near 25~notes/s. Three outcomes are possible. The perceptual inflection may coincide with the measured separability peak, confirming the hypothesis. It may occur at a different density, suggesting that the metric captures distributional structure that does not map directly onto perceptual categorisation. Or no clear sigmoidal transition may appear, indicating that the melodic-to-textural shift is not well described by a single density threshold for piano timbres.

A second task addresses pipeline-fidelity perception. Participants hear the 74-second canonical composition (Excerpt~3) and press a button whenever they perceive a ``change in musical character''. The temporal distribution of button presses is compared against the actual symbol-transition timestamps using a permutation test. If the mean absolute deviation between button presses and symbol boundaries is significantly smaller than chance (estimated by circularly shifting press times), this would confirm that pipeline-level distributional changes are perceptually salient.

Power analysis (Cohen's $d = 0.8$, $\alpha = .05$, power $= .80$) confirms the suitability of the proposed sample size ($N = 30$). This protocol is designed to be executable with existing additional materials (audio) and standard psychoacoustic software (e.g.\ PsychoPy).

A listener study answers only half of the question this system raises. Amanous is proposed as an augmented compositional instrument (Section~\ref{sec:bg_creativity}), and an instrument is evaluated not only by what an audience hears but by what a composer can do with it. That second question calls for a qualitative study of the compositional experience itself, in which composers work with the system over an extended period and their practice is documented through think-aloud protocols during parameter tuning, semi-structured interviews, and analysis of the revision histories left behind by successive parameter settings. Such a study would ask which parts of the control surface afford compositional intent and which resist it, whether the information-theoretic feedback (PCC, VSS) is legible as a compositional signal rather than a diagnostic readout, and whether distribution-switching is experienced as a musical decision or as a technical configuration step. The answers bear directly on the design of the system, since a parameter that no composer can form an intention about is a parameter in the wrong place. This line of work is complementary to the psychoacoustic protocol above and is left to future research.

\section{Conclusion}
\label{sec:conclusion}

This paper asked whether compositional intent can be transmitted from grammar symbol to actuation-ready MIDI through a unified pipeline and what measurable constraints govern each stage of that transmission. Amanous demonstrates that this is achievable by unifying tempo canons, stochastic distributions, and L-system grammars via distribution-switching within a four-layer architecture.

\begin{enumerate}
    \item \textbf{Integration architecture.} Distribution-switching produces statistically distinct musical sections ($d = 9.67$ for melodic and $d = 8.74$ for rhythmic coherence). Degradation analysis identifies Layer~4 as the one stage that perturbs the intended distributions, through latency compensation where velocity varies and through the key-reset mask where a key is re-struck within 50~ms, while ablation shows the non-redundant contribution of the tempo canon and the hardware layer, and locates the L-system's contribution at the macro-formal level, significantly so only at greater expansion depths.

    \item \textbf{Hardware abstraction layer.} Layer~4 pre-compensates velocity-dependent latency and enforces the key-reset constraint on every emitted trigger, at a measured cost of 8.5\% of the events in the canonical piece. Both rest on a stated model of the instrument rather than on measurement of one.

    \item \textbf{Operational envelope.} Structured and random textures are most separable near 25~notes/s, and single-domain melodic metrics lose that separability over the 40--100~notes/s band. Beyond it, dynamic stratification, not melodic coherence, becomes the primary lever for inter-voice differentiation (98.3\% of nwVSS), and what separates structured from random material is consistent with pitch-class content rather than melodic order.

    \item \textbf{Convergence point calculus.} CP events provide a deterministic-to-stochastic control interface, verified for discrete triggering and for continuous rate tracking ($r = .928$), with $\epsilon$ offering monotonic convergence-resolution control.
\end{enumerate}

All reported results are computational. The principal limitations are the absence of perceptual validation and platform specificity to the Yamaha Disklavier. Future work should test the measured separability peak and loss band against perceptual thresholds through the proposed psychoacoustic protocol (Section~\ref{sec:future_protocol}), extend the latency model with instrument-specific calibration data, and explore adaptation to other automated instruments.

As an augmented compositional instrument, Amanous enables reproducible, principled navigation of parameter spaces that exceed human physical capacity, bridging the gap between algorithmic abstraction and electromechanical reality.

Supplementary materials accompany this article as a compressed audio archive. Source code: \url{https://github.com/joonhyungbae/amanous}.

\printbibliography

\clearpage
\appendix

\section{Complete Statistical Summary}
\label{app:stats}

\begin{table}[htbp]
\tbl{Complete statistical results summary. All tests two-tailed, $\alpha = .05$.}
{%
\resizebox{\linewidth}{!}{%
\begin{tabular}{@{}p{3.5cm}lllll@{}} \toprule
\textbf{Comparison} & \textbf{$N$} & \textbf{Test} & \textbf{Statistic} & \textbf{$p$} & \textbf{Effect [CI]} \\ \midrule
\multicolumn{6}{@{}l}{\textit{Pipeline Fidelity (Section~\ref{sec:results_ds})}} \\
Melodic coherence & 28 & $t$ & $t(26)=25.53$ & $<.001$ & $d=9.67$ $[7.03, 12.31]$ \\
Rhythmic coherence & 28 & $t$ & $t(26)=23.07$ & $<.001$ & $d=8.74$ $[6.33, 11.15]$ \\
Pitch entropy (PCC) & 8 & M-W & $U=15.0$ & $.036$ & $d=75.98$ \\
\midrule
\multicolumn{6}{@{}l}{\textit{Component Ablation (Section~\ref{sec:ablation})}} \\
(a) Seq.\ self-sim.\ MC & 100\textsuperscript{a} & $z$ & $z=-0.40$ & $.692$ & n/a \\
(a) Same-symbol MC & 100\textsuperscript{a} & $z$ & $z=-1.83$ & $.068$ & n/a \\
(a) Same-symbol RC & 100\textsuperscript{a} & $z$ & $z=-1.01$ & $.312$ & n/a \\
(b) VSS temporal & 8+8 & M-W & $U=64.0$ & $<.001$ & $r=-1.00$; $-78.4\%$ \\
(b) RC same-symbol & 8+8 & M-W & $U=51.0$ & $.050$ & $r=-.59$; $-18.1\%$ \\
(c) Vel.--timing $r$ (lin.) & n/a & n/a & $r=-1.000$ & n/a & n/a \\
(c) Vel.--timing $r$ (pow.) & n/a & n/a & $r=-.998$ & n/a & n/a \\
\midrule
\multicolumn{6}{@{}l}{\textit{Hardware Simulation (Section~\ref{sec:hardware})}} \\
Vel.-error (uncorr.) & 526 & Pearson & $r=-.971$ & $<.001$ & n/a \\
Filter error SD & 526 & paired $t$ & n/a & $<.001$ & $-32.4\%$ \\
\midrule
\multicolumn{6}{@{}l}{\textit{Coherence Metrics (Section~\ref{sec:coherence})}} \\
Peak separability & 14\textsuperscript{b} & bootstrap & \multicolumn{2}{l}{25 notes/s [CI: 25, 25]} & gap $d = 8.06$ \\
Constraint PCC & 90 & K-W & $H=61.60$ & $<.001$ & n/a \\
Multi-constr.\ VSS & 18\textsuperscript{c} & K-W & $H=12.32$ & $.002$ & $+1849\%$ \\
Pitch-vel coupling\textsuperscript{d} & 6+6\textsuperscript{c} & M-W & $U=0.0$ & $.002$ & $+10{,}503\%$ \\
\midrule
\multicolumn{6}{@{}l}{\textit{Convergence Points (Section~\ref{sec:convergence})}} \\
CP density switch & 30\textsuperscript{e} & M-W & $U=0$ & $<.001$ & $r=1.00$ \\
CP PCC switch & 30\textsuperscript{e} & M-W & $U=900$ & $<.001$ & $r=-1.00$ \\
Continuous tracking & 30 & Pearson & $r=.928$ & $<.001$ & n/a \\ \bottomrule
\end{tabular}%
}%
}
\tabnote{M-W: Mann-Whitney $U$; K-W: Kruskal-Wallis $H$.\\ \textsuperscript{a}$N = 100$ random permutations; $z = (\text{full} - \mu_{\text{random}}) / \sigma_{\text{random}}$, $p$ from standard normal.\\ \textsuperscript{b}$N = 14$ density levels; the peak-separability CI is bootstrapped over 40 trials per level.\\ \textsuperscript{c}Voice pairs per condition (four voices give six pairs).\\ \textsuperscript{d}The pitch-velocity coupling makes velocity a deterministic function of pitch, so the near-unit correlation and the wVSS increase verify that the constraint was applied rather than reporting an independent effect.\\ \textsuperscript{e}$N$ reports one-second analysis windows of the Poisson voice (30 pre-CP and 30 post-CP for the discrete switch, from a 60-second run).}
\label{tab:stats_full}
\end{table}

\section{Constraint Inequality Summary}
\label{app:constraints}

Table~\ref{tab:constraint_ineq} collects the inequalities the pipeline enforces at Layer~4, together with the two metric-derived boundaries reported in this study.

\begin{table}[htbp]
\tbl{Complete constraint system.}
{\begin{tabular}{@{}p{3.5cm}lp{4cm}@{}} \toprule
\textbf{Constraint} & \textbf{Inequality} & \textbf{Source} \\ \midrule
\multicolumn{3}{@{}l}{\textit{Hardware}} \\
Velocity range & $v \in [0, 1023]$ & 10-bit XPMIDI \\
Per-key rate & $\text{IOI}_{\text{key}} \geq 50$~ms & Key reset time \\
Latency bounds & $10 \leq L(v) \leq 30$~ms & Modelling assumption\textsuperscript{a} \\
Polyphony & $N_{\text{sim}} \leq 88$ & Physical keys \\
Scanning resolution & $\Delta t_{\min} \approx 1$~ms & 800--1000~Hz \\ \midrule
\multicolumn{3}{@{}l}{\textit{Metric-derived boundaries}} \\
Peak separability & $\rho_{\text{agg}} \approx 25$~notes/s\textsuperscript{b} & This study \\
Registral separation & $\Delta p > 5$ semitones & This study (heuristic)\textsuperscript{c} \\ \bottomrule
\end{tabular}}
\tabnote{\textsuperscript{a}No published velocity-to-onset-latency calibration exists for any Disklavier model (Appendix~\ref{app:audit}). The bounds are the assumed envelope of $L(v)$, to be replaced by measured values per instrument (Section~\ref{sec:hardware}).\\
\textsuperscript{b}Derived from stochastic two-voice textures, with per-voice density $\approx \rho_{\text{agg}}/2$. Perceptual correspondence requires experimental validation.\\
\textsuperscript{c}An engineering heuristic of this study, not a perceptual threshold. \citet{huron2001tone} gives a pitch-proximity figure of roughly one to two semitones, but it bounds the interval between successive tones \emph{within} a voice, not the gap between concurrent voices.}
\label{tab:constraint_ineq}
\end{table}

\FloatBarrier
\section{\texorpdfstring{Power-Law Exponent $c$ Sensitivity (Layer~4)}{Power-Law Exponent c Sensitivity (Layer 4)}}
\label{app:powerlaw_sensitivity}

To confirm that the compensation behaviour of Layer~4 is not tied to a single choice of power-law exponent $c$, we varied $c \in \{0.3, 0.4, 0.5, 0.6, 0.7\}$ and computed the standard deviation of the residual jitter (ms) after compensation, assuming that the true latency model has $c = 0.5$. The residual error per note is $L_{\text{true}}(v) - L_c(v)$ in ms over the same 526-note excerpt. The residual jitter is zero when the compensation exponent matches the true model ($c = 0.5$) and increases smoothly as $c$ deviates (Table~\ref{tab:sensitivity_powerlaw_c}), showing a consistent trend rather than dependence on a single parameter. 

\begin{table}[htbp]
\tbl{Residual jitter standard deviation (ms) after Layer~4 compensation for varying power-law exponent $c$ ($N = 526$ notes, true $c = 0.5$).}
{\begin{tabular}{@{}lc@{}} \toprule
$c$ & \textbf{Residual Jitter SD (ms)} \\ \midrule
0.3 & 1.0446 \\
0.4 & 0.4769 \\
0.5 & 0.0000 \\
0.6 & 0.4015 \\
0.7 & 0.7404 \\ \bottomrule
\end{tabular}}

\label{tab:sensitivity_powerlaw_c}
\end{table}

\section{Latency Model Mismatch (HAL vs.\ No Correction)}
\label{app:latency_mismatch}

To test whether HAL compensation remains beneficial when the true piano latency deviates from the assumed model, we ran two mismatch simulations on the same 526-note excerpt. In the first, an exponent mismatch, true latency follows a power-law with $c_{\mathrm{true}} \in [0.3, 0.35, \ldots, 0.7]$ while HAL uses $c = 0.5$. The uncorrected jitter is the standard deviation of $L_{\mathrm{true}}(v)$, and the jitter with HAL is the standard deviation of $L_{\mathrm{true}}(v) - L_{0.5}(v)$. In the second, additive noise, $L_{\mathrm{actual}}(v) = L(v; c=0.5) + U(-w, +w)$~ms with $w \in \{0, 0.5, 1, 1.5, 2\}$ and 200 trials per $w$. In every condition, HAL reduces jitter relative to the uncorrected baseline (Table~\ref{tab:mismatch_summary}), so the sensitivity curves show that correction is always preferable to no correction despite model error. 

To test the claim that the power-law assumption alone yields benefit without perfect calibration, we added a three-condition simulation. A ``virtual real piano'' was defined as the nominal power-law ($c = 0.5$, 10--30~ms) with $\pm 10\%$ per-note multiplicative noise and parameter drift (exponent $c$ varying by note index via a bounded random walk). For each of 200 trials, we computed the onset-timing jitter (std, ms) under three conditions, namely (A) no correction (raw MIDI), (B) Amanous HAL with $c = 0.5$, and (C) ideal correction, in which the compensation matches the virtual model exactly. Condition (C) is a reference bound rather than an experimental result, since a perfectly matched correction has zero error by construction.

Table~\ref{tab:virtual_real_piano} reports the jitter under each condition. Jitter with HAL (B) was significantly lower than without correction (A), by a paired $t$-test ($p < .001$, mean difference of $2.70$~ms, Cohen's $d = 29.04$), and the two conditions do not overlap across the 200 trials. The effect size, rather than the $p$ value, is what carries this claim, since the number of simulated trials is a free parameter and the $p$ value can be driven arbitrarily low by running more of them. Even when the true latency deviates from the assumed model through noise and drift, following the general physical law alone provides a substantial improvement over raw MIDI.

\begin{table}[htbp]
\tbl{Latency model mismatch: jitter (ms, SD) with HAL applied vs.\ uncorrected.}
{\begin{tabular}{@{}lrr@{\qquad}lrr@{}} \toprule
\multicolumn{3}{c}{\textbf{Exponent mismatch}} & \multicolumn{3}{c}{\textbf{Additive noise $\pm w$~ms}} \\
\cmidrule(r){1-3}\cmidrule(l){4-6}
$c_{\mathrm{true}}$ & \textbf{Uncorr.} & \textbf{HAL} & $w$ & \textbf{Uncorr.} & \textbf{HAL} \\ \midrule
0.30 & 2.47 & 1.04 & 0 & 3.50 & 0.00 \\
0.50 & 3.50 & 0.00 & 1.0 & 3.55 & 0.58 \\
0.70 & 4.21 & 0.74 & 2.0 & 3.69 & 1.15 \\ \bottomrule
\end{tabular}}
\tabnote{Uncorr.\ = uncorrected (no HAL). HAL = compensation with $c=0.5$. $N=526$ notes. For the noise case, values are means over 200 trials.}
\label{tab:mismatch_summary}
\end{table}

\begin{table}[htbp]
\tbl{Virtual real piano simulation. Onset jitter in ms (mean $\pm$ std over 200 trials, $N = 526$ notes per trial).}
{\begin{tabular}{@{}lc@{}}
\toprule
\textbf{Condition} & \textbf{Jitter (mean $\pm$ std, ms)} \\
\midrule
(A) Raw (no correction) & $3.63 \pm 0.10$ \\
(B) Amanous HAL ($c=0.5$) & $0.93 \pm 0.03$ \\
(C) Ideal & $0.00$ \\
\bottomrule
\end{tabular}}
\label{tab:virtual_real_piano}
\end{table}

\section{\texorpdfstring{$\epsilon$ Sensitivity in CP Calculus}{Epsilon Sensitivity in CP Calculus}}
\label{app:epsilon_sensitivity}

\begin{figure}[htbp]
\centering
\includegraphics[width=0.95\textwidth]{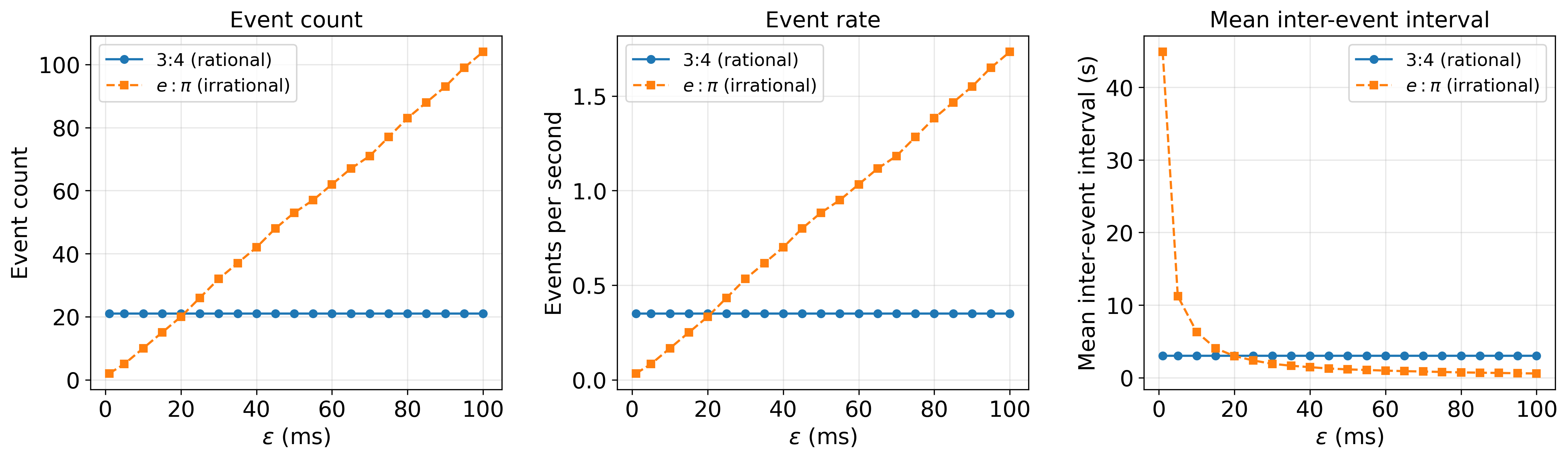}
\caption{$\epsilon$ sensitivity over a 60~s sweep, showing event count, event rate, and mean inter-event interval for the rational 3:4 canon (solid line, circles) and the irrational $e{:}\pi$ canon (dashed line, squares).}
\label{fig:epsilon_sensitivity_full}
\end{figure}

To confirm that the CP switching behaviour is not tied to a single choice of $\epsilon$ and to demonstrate that $\epsilon$ acts as a compositional parameter for \emph{texture transition density}, we ran two analyses. The first is a coarse grid over 30~s with $\epsilon \in \{10, 20, 50, 100\}$~ms (Table~\ref{tab:epsilon_sensitivity}). The second is a full sweep over 60~s with $\epsilon = 1, 5, 10, \ldots, 100$~ms (5~ms step), and for each $\epsilon$ we computed the number of convergence events and the inter-event intervals (Definition~\ref{def:cp}). The results are shown in Figure~\ref{fig:epsilon_sensitivity_full}.

For the rational 3:4 canon (IOI 1.0~s and 0.75~s), the event count is invariant across $\epsilon$ (exact convergences every 3~s; 11 events in 30~s, 21 in 60~s). For the irrational canon $e:\pi$, the count increases monotonically with $\epsilon$. In the full sweep, the correlation between $\epsilon$ and the event count is $r \approx .9998$, and the mean interval between events decreases from $\approx 45$~s at $\epsilon = 1$~ms to $\approx 0.58$~s at $\epsilon = 100$~ms. Thus, $\epsilon$ is not merely an error tolerance, but a continuous control over the frequency with which distribution switches occur.

\begin{table}[htbp]
\tbl{Convergence-point count and switching frequency for varying $\epsilon$ ($N = 30$~s).}
{\begin{tabular}{@{}lrrrr@{}} \toprule
$\epsilon$ (ms) & \textbf{3:4 count} & \textbf{3:4 (/s)} & \textbf{$e:\pi$ count} & \textbf{$e:\pi$ (/s)} \\ \midrule
10 & 11 & 0.367 & 5 & 0.167 \\
20 & 11 & 0.367 & 11 & 0.367 \\
50 & 11 & 0.367 & 26 & 0.867 \\
100 & 11 & 0.367 & 51 & 1.700 \\ \bottomrule
\end{tabular}}
\tabnote{Rational 3:4: count stable across $\epsilon$. Irrational $e:\pi$: count increases with $\epsilon$.}
\label{tab:epsilon_sensitivity}
\end{table}

\section{nwVSS Weights by Density}
\label{app:nwvss_density}

Applying the same nwVSS weight-extraction procedure (Section~\ref{sec:wvss_validation}) to low-density conditions (20~notes/s aggregate) and high-density (120~notes/s aggregate) yields the weights in Table~\ref{tab:nwvss_density}. The weight shift from high to low density is $-7.81$ percentage points for velocity, $+7.63$ for temporal, and $+0.18$ for pitch. At low density, velocity dominance weakens and temporal (and pitch) components gain relative weight (\emph{weight transfer by density}). The 0.00\% temporal weight at high density is a computational signature consistent with the aggregate-texture account, although the metric measures inter-voice distance and not perception. At 120~notes/s, the onset structure ceases to carry discriminative load for separation, which the metric correctly attributes to velocity and pitch.

\begin{table}[htbp]
\tbl{nwVSS weights by aggregate density (same extraction procedure).}
{\begin{tabular}{@{}lrrr@{}} \toprule
Condition & $w_{\text{pitch}}$ (\%) & $w_{\text{vel}}$ (\%) & $w_{\text{temporal}}$ (\%) \\ \midrule
Low density (20~notes/s) & 0.45 & 91.92 & 7.63 \\
High density (120~notes/s) & 0.27 & 99.73 & 0.00 \\ \bottomrule
\end{tabular}}
\tabnote{Weight shift (low minus high): $\Delta w_{\text{pitch}} = +0.18$, $\Delta w_{\text{vel}} = -7.81$, $\Delta w_{\text{temporal}} = +7.63$ percentage points.}
\label{tab:nwvss_density}
\end{table}

\section{Calibration Data Set Audit}
\label{app:audit}

Six publicly available Disklavier-related repositories were audited for velocity-latency calibration data: MAPS Database (audio-MIDI pairs, no velocity sweep), Magenta MIDI Dataset (no hardware-specific latency), MAESTRO (no Disklavier-specific latency characterisation), SMD (score-performance alignment only), Vienna 4$\times$22 Piano Corpus (tone quality focus), and RWC Music Database (no systematic latency measurement). No public data set pairs MIDI velocity commands with measured acoustic-onset latency for any Disklavier model. Controlled velocity sweeps ($v = 0, 1, \ldots, 1023$; multiple repetitions) with sub-millisecond audio capture are recommended.

\section{Hardware-Aware Actuation Pipeline}
\label{app:algorithm}

Algorithm~\ref{alg:latency} presents the complete adaptive compensation pipeline, which replaces the simplified correction in Algorithm~\ref{alg:generation} (Line~11).

\begin{algorithm}[H]
\caption{Hardware-Aware Actuation Pipeline (HAL)}
\label{alg:latency}
\begin{algorithmic}[1]
\Require Event list $E$, calibration data $C$ (optional), compression $\gamma$
\Ensure Compensated event list $E'$
\If{$C$ is available}
    \State $f \gets \textsc{FitPowerLaw}(C)$ \Comment{Prior $c = 0.5$ until refitted}
    \State $L(v) \gets 30 - 20 \cdot f(v/1023)$
    \State \textbf{skip} robustness filter
\Else
    \State $L(v) \gets 30 - 20 \cdot (v/1023)$ \Comment{Linear fallback}
\EndIf
\State $E' \gets \emptyset$
\For{each event $(t, p, v, d)$ in $E$}
    \If{$C$ \textbf{not} available \textbf{and} \textsc{IsLatencySensitive}$(E, t, p)$}
        \State $\bar{v} \gets$ mean velocity in $[t-25\text{ms}, t+25\text{ms}]$
        \State $v \gets \bar{v} + \gamma \cdot (v - \bar{v})$ \Comment{$\gamma < 1$: compress}
    \EndIf
    \State $t' \gets t - L(v)/1000$
    \State $E' \gets E' \cup \{(t', p, v, d)\}$
\EndFor
\State \Return $E'$
\end{algorithmic}
\end{algorithm}

\section{Supplementary Materials}
\label{app:supplementary}

Supplementary materials (audio of compositions generated algorithmically by the framework) accompany this article as a compressed archive. These materials document the algorithmically generated output and are for reference and future perceptual experiments. The quantitative results in this paper do not rely on empirical measurement of these recordings. Source code: \url{https://github.com/joonhyungbae/amanous}.

The following excerpts are referenced in the text and match the materials in the supplementary archive:
\begin{itemize}
    \item \textbf{Excerpt 1:} Demonstration of beyond-human density, rendered from MIDI generated for the Disklavier (34~s). Polyphony (40-note chords), 30~Hz multi-key trill, 6-octave arpeggio.
    \item \textbf{Excerpt 2:} Phase Music, a Minimalist Study (80~s). Reich-inspired phase-shift, with a pentatonic set and a 1:1.01 tempo drift.
    \item \textbf{Excerpt 3:} Canonical $ABAABABA$ validation composition (74~s). L-system macro-form with deterministic sections ($A$) and textural sections ($B$), with a 3:4 tempo canon.
    \item \textbf{Excerpt 4:} Convergence Point demonstration, 3:4 canon (30~s). Pre-CP sparse/melodic and post-CP dense texture switch at $t = 15$~s.
\end{itemize}

The excerpts are software renders of the generated MIDI through a sampled piano, not recordings of a Disklavier, and they can also be heard at \url{https://joonhyungbae.github.io/amanous/}. The code repository additionally holds (1)~the MIDI files and event lists of the excerpts, (2)~the ablation scripts, with deterministic seeds and JSON/CSV output, and (3)~the analysis scripts and saved outputs from which the reported statistics are computed, indexed by table in the repository's README.

\end{document}